\documentclass[aps,prb,twocolumn,amsmath,amssymb,groupedaddress,longbibliography
]{revtex4-2}

\usepackage[dvipdfmx]{graphicx}
\usepackage{dcolumn}
\usepackage{bm}
\usepackage{multirow}
\usepackage{braket}
\usepackage{color}
\usepackage{ulem}
\usepackage{longtable}
\usepackage{booktabs}

\allowdisplaybreaks[4]

\begin{document}

\title{
Unconventional Hall effect and magnetoresistance induced by metallic ferroaxial ordering
}

\author{Satoru Hayami$^{1}$, Rikuto Oiwa$^2$, and Hiroaki Kusunose$^{3,4}$}
\affiliation{
$^1$Graduate School of Science, Hokkaido University, Sapporo 060-0810, Japan \\
$^2$RIKEN Center for Emergent Matter Science, Wako, Saitama 351-0198, Japan \\
$^3$Department of Physics, Meiji University, Kawasaki 214-8571, Japan \\
$^4$Quantum Research Center for Chirality, Institute for Molecular Science, Okazaki 444-8585, Japan
}

\begin{abstract}
Transport property under metallic ferroaxial ordering in an external magnetic field is theoretically investigated. 
After presenting the relation between the magnetoconductivity tensor and ferroaxial moment from the symmetry viewpoint, we analyze the behavior of the unconventional Hall effect and magnetoconductivity for a general five $d$-orbital tight-binding model under the point group $C_{\rm 4h}$, where the ferroaxial moment is activated. 
We show that the crystalline electric field that arises from the symmetry reduction from $D_{\rm 4h}$ to $C_{\rm 4h}$ is essential for the ferroaxial-related magnetotransport, while the relativistic spin--orbit coupling is not required. 
We also compare the unconventional Hall effect driven by the ferroaxial moment with the conventional Hall effect, the latter of which does not require the ferroaxial moment. 
The present results provide characteristic transport properties in the ferroaxial systems, which can be observed in various candidate materials like Ca$_5$Ir$_3$O$_{12}$. 
\end{abstract}

\maketitle

\section{Introduction}

Ferroaxial order has attracted much attention as a new quantum state of matter, which is qualitatively distinct from ferromagnetic, ferroelectric, and ferro-magnetoelectric (magnetic toroidal) orders in terms of spatial inversion and time-reversal parities~\cite{Hlinka_PhysRevLett.116.177602,cheong2018broken, cheong2022linking}. 
Its order parameter is characterized by a time-reversal-even axial vector, which means that the ferroaxial moment can be activated even in the paramagnetic state with centrosymmetric lattice structures. 
Such ferroaxial order has been mainly studied in dielectric materials, where the vortex structure of the electric dipole moments corresponding to a signature of the ferroaxial moment emerges in nanodisks and nanorods~\cite{naumov2004unusual, Ponomareva_PhysRevB.72.214118,prosandeev2008original, Naumov_PhysRevLett.101.107601, Stachiotti_PhysRevLett.106.137601}. 
So far, several materials have been identified to be ferroaxial ordered states in bulk systems, such as Co$_3$Nb$_2$O$_8$~\cite{Johnson_PhysRevLett.107.137205}, CaMn$_7$O$_{12}$~\cite{Johnson_PhysRevLett.108.067201}, RbFe(MoO$_4$)$_2$~\cite{jin2020observation,Hayashida_PhysRevMaterials.5.124409}, NiTiO$_3$~\cite{hayashida2020visualization, Hayashida_PhysRevMaterials.5.124409, yokota2022three, Guo_PhysRevB.107.L180102}, Ca$_5$Ir$_3$O$_{12}$~\cite{Hasegawa_doi:10.7566/JPSJ.89.054602, hanate2021first, hayami2023cluster, hanate2023space}, BaCoSiO$_4$~\cite{Xu_PhysRevB.105.184407}, K$_2$Zr(PO$_4$)$_2$~\cite{yamagishi2023ferroaxial}, Na$_2$Hf(BO$_3$)$_2$~\cite{nagai2023chemicalSwitching}, and Na-superionic conductors~\cite{nagai2023chemical}. 

The microscopic origin of ferroaxial ordering is described by odd-rank electric toroidal multipoles, such as the electric toroidal dipole and octupole, whose expression at the classical level is obtained by performing the multipole expansion under the assumption that the magnetic charge (monopole) exists in Maxwell's equation~\cite{dubovik1975multipole,dubovik1986axial,dubovik1990toroid,hayami2018microscopic}. 
In this case, the electric toroidal dipole is described by the outer product of the position vector and the electric dipole. 
Meanwhile, recent theoretical studies revealed an atomic-scale description of the electric toroidal dipole at the quantum-mechanical level by introducing the symmetry-adapted multipole basis~\cite{kusunose2020complete}, where it can be also described by the outer product of the orbital and spin angular momenta~\cite{Hayami_PhysRevB.98.165110}. 
Such fundamental progress in terms of the microscopic description of the electric toroidal dipole provides the possibility of the atomic-scale ferroaxial order and a way of the quantitative evaluation based on the ab-initio calculations~\cite{hoshino2022spin}.

In parallel with the development of microscopic expressions, physical phenomena driven by the ferroaxial moment have been extensively investigated~\cite{cheong2021permutable, Nasu_PhysRevB.105.245125, Roy_PhysRevMaterials.6.045004, Hayami_PhysRevB.106.144402, Hayami_doi:10.7566/JPSJ.91.113702, cheong2022linking, inda2023nonlinear}. 
As the ferroaxial moment breaks the mirror symmetry parallel to its direction, it acts as a rotator against external stimuli~\cite{Hayami_doi:10.7566/JPSJ.91.113702}. 
For example, when the ferroaxial moment lies in the $z$ direction, an input field along the $x$ ($y$) direction induces the response of the conjugate physical quantities along the $y$ ($-x$) direction. 
In other words, the ferroaxial moment becomes a source of transverse responses between the input and output fields/currents with the same symmetry. 
These intriguing features would open a new research field with the use of the ferroaxial moment, i.e., ''ferroaxiality", as with ferromagnetism and ferroelectricity. 

In the present study, we further investigate physical phenomena under metallic ferroaxial ordering. 
We especially focus on two transport phenomena under an external magnetic field: One is the unconventional Hall effect and the other is the magnetoconductivity, both of which are caused by a ferroaxial moment as nanometric rotator. 
The former is characterized by the off-diagonal antisymmetric part of the linear conductivity tensor proportional to an external magnetic field $H$, while the latter is by its symmetric part proportional to $H^2$. 
We perform symmetry and model analyses for a general five $d$-orbital model consisting of electron hopping, relativistic spin--orbit coupling, and tetragonal crystalline electric field under the point group $C_{\rm 4h}$. 
As a result, we find that the crystalline electric field arising from the symmetry reduction from $D_{\rm 4h}$ to $C_{\rm 4h}$ leads to the unconventional Hall effect and magnetoconductivity characteristic of the metallic ferroaxial moment. 
Our results indicate that the ferroaxial system exhibits characteristic transport properties, which would be observed in ferroaxial materials, such as Ca$_5$Ir$_3$O$_{12}$~\cite{Matsuhira_doi:10.7566/JPSJ.87.013703}.

The rest of the paper is organized as follows. 
In Sec.~\ref{sec: Setup}, we present the tight-binding model in the tetragonal lattice structure. 
We also introduce the ferroaxial moments described by the electric toroidal dipole, octupole, and electric hexadecapole. 
Among them, we discuss the behavior of the field-induced electric toroidal dipole. 
We give symmetry analysis and numerical results in terms of the magnetoconductivity tensor under an external magnetic field in Secs.~\ref{sec: Planar Hall effect} and \ref{sec: Magnetoconductivity}. 
In Sec.~\ref{sec: Planar Hall effect}, we discuss the behavior of the unconventional Hall conductivity with an odd function of $H$, and in Sec.~\ref{sec: Magnetoconductivity}, we discuss that of the magnetoconductivity with an even function of $H$.  
Section~\ref{sec: Summary} is devoted to a summary of this paper. 
We present the essential model parameters for the field-induced electric toroidal dipole, unconventional Hall conductivity, and magnetoconductivity in Appendix~\ref{sec: Essential model parameters}. 
In Appendix~\ref{sec: Optical conductivity}, we show the result of the optical unconventional Hall conductivity.

\section{Setup}
\label{sec: Setup}

In this section, we introduce the tight-binding model in the three-dimensional tetragonal lattice system under the point group $C_{\rm 4h}$ in Sec.~\ref{sec: Model}. 
Then, we discuss the behavior of the atomic-scale electric toroidal dipole induced by the external magnetic field in Sec.~\ref{sec: Ferroaxial moment}. 

\subsection{Model}
\label{sec: Model}

\begin{figure}[htb!]
\begin{center}
\includegraphics[width=1.0 \hsize]{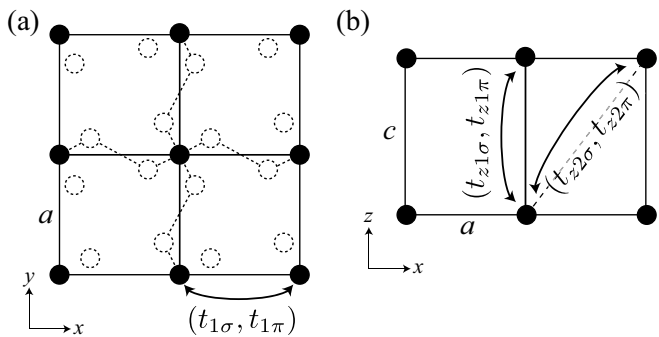}
\caption{
\label{Fig: lattice}
The lattice structure with the lattice constants $a$ and $c$ under the point group $C_{\rm 4h}$ viewed from (a) the $z$ axis and (b) the $y$ axis. 
The black-filled circles represent the $d$-orbital sites, while the black-dashed circles represent the ligand sites.
The hopping parameters, $(t_{1\sigma},t_{1\pi}, t_{z1\sigma}, t_{z1\pi}, t_{z2\sigma}, t_{z2\pi})$, are presented. 
}
\end{center}
\end{figure}

We consider the tight-binding model consisting of five $d$ orbitals, where the lattice structure is set as a three-dimensional square-lattice structure belonging to the point group $C_{\rm 4h}$; the vertical mirror plane perpendicular to the $xy$ plane is broken. 
We here suppose that such a lack of vertical mirror symmetry is caused by the ligand ions (the black-dashed circles) around the $d$-orbital sites (the black-filled circles), as shown in Fig.~\ref{Fig: lattice}(a); the point group is $D_{\rm 4h}$ if the ligand ions are absent. 
The square layers in Fig.~\ref{Fig: lattice}(a) are stacked along the $z$ direction, as shown in Fig.~\ref{Fig: lattice}(b). 
We set the lattice constants as $a=c=1$ for simplicity.
Then, the Hamiltonian is given by
\begin{align}
\label{eq: Ham}
\mathcal{H}&=\mathcal{H}_t + \mathcal{H}_{\rm SOC} +  \mathcal{H}_{\rm CEF}+\mathcal{H}_{V}+\mathcal{H}_{\rm Z},\\
\label{eq: Hamt}
\mathcal{H}_t&= \sum_{ij \alpha\beta\sigma}(t_{\alpha\beta}^{ij}  c^{\dagger}_{i \alpha \sigma}c_{j \beta\sigma}+{\rm H.c.}),\\
\label{eq: HamSOC}
\mathcal{H}_{\rm SOC}&= \lambda \sum_i \bm{l}_i\cdot \bm{s}_i,  \\
\label{eq: HamCEF}
\mathcal{H}_{\rm CEF}&= \sum_{i\sigma} (
\Delta_1 c^{\dagger}_{i xy \sigma}c_{i xy \sigma}
+\Delta_2 c^{\dagger}_{i v \sigma}c_{i v \sigma}
+\Delta_3 c^{\dagger}_{i u \sigma}c_{i u \sigma}
),\\
\label{eq: HamV}
\mathcal{H}_{V}&= V \sum_{i \sigma}(c^{\dagger}_{i xy \sigma}c_{i v \sigma}+c^{\dagger}_{i  v \sigma}c_{i xy \sigma}),\\
\label{eq: HamZeeman}
\mathcal{H}_{\rm Z}&=-\sum_{i} \bm{H} \cdot  (\bm{l}_i + 2 \bm{s}_i)
\end{align}
where $c^\dagger_{i\alpha\sigma}$ ($c_{i\alpha \sigma}$) is the creation (annihilation) operator of an electron with site $i$, orbital $\alpha=(u=3z^2-r^2, v=x^2-y^2, yz, zx, xy)$ for five $d$ orbitals ($d_{u}, d_{v}, d_{yz}, d_{zx}, d_{xy}$), and spin $\sigma$.

The first term in Eq.~(\ref{eq: Ham}) represents the hopping Hamiltonian $\mathcal{H}_t$ in Eq.~(\ref{eq: Hamt}). 
We consider nonzero hopping elements allowed from the symmetry based on the Slater-Koster table; we parametrize the nearest-neighbor hopping in the $xy$ plane, $(t_{1\sigma} ,t_{1\pi})$ [Fig.~\ref{Fig: lattice}(a)], and the nearest-neighbor and next-nearest-neighbor hoppings along the $z$ direction, $(t_{z1\sigma},t_{z1\pi})$ and $(t_{z2\sigma},t_{z2\pi})$ [Fig.~\ref{Fig: lattice}(b)], where we ignore the $\delta$ component of the Slater-Koster parameters for simplicity. 
We set $t_{1\sigma}=-1$, $t_{1\pi}=0.5$, $t_{z1\sigma}=-0.8$, $t_{z1\pi}=0.4$, $t_{z2\sigma}=-0.5$, and $t_{z2\pi}=0.25$; $t_{1\sigma}$ is the energy unit of the model. 
The second term in Eq.~(\ref{eq: Ham}) represents the spin--orbit coupling Hamiltonian $\mathcal{H}_{\rm SOC}$ in Eq.~(\ref{eq: HamSOC}), where $\bm{l}_i$ and $\bm{s}_i$ represent the orbital and spin angular momentum operators, respectively; we set $\lambda=2$ unless otherwise stated by implicitly considering the 5$d$ materials where $\lambda$ tends to be comparable to the hopping parameters~\cite{Charlebois_PhysRevB.104.075153}.  
The third term in Eq.~(\ref{eq: Ham}) represents the Hamiltonian for the crystalline electric field (CEF) under the point group $D_{4\rm h}$ $\mathcal{H}_{\rm CEF}$ in Eq.~(\ref{eq: HamCEF}).
We take $\Delta_1=0.5 b$, $\Delta_2=6.5 b$, and $\Delta_3=10.5 b$ with $b=\sqrt{3}/2$, where the atomic-orbital energy levels $E_{\alpha}$ satisfy $E_{yz}=E_{zx}<E_{xy}<E_{v}<E_{u}$. 
The fourth term in Eq.~(\ref{eq: Ham}) represents the additional CEF Hamiltonian in Eq.~(\ref{eq: HamV}) to represent the symmetry lowering from $D_{4\rm h}$ to $C_{\rm 4h}$. 
The CEF parameter $V$ mixes the $xy$ and $x^2-y^2$ orbitals; we set $V=0.7$.  
The choice of the model parameters does not alter qualitative features in the following sections.
In particular, a different ratio of the spin--orbit coupling and CEF as found in 3$d$, 4$d$, and 5$d$ materials does not affect the following results at the qualitative level; we demonstrate this argument by performing the method to extract the essential model parameters in response tensors in Appendix~\ref{sec: Essential model parameters}.  
The last term in Eq.~(\ref{eq: Ham}) represents the Zeeman Hamiltonian in Eq.~(\ref{eq: HamZeeman}) for an external magnetic field $\bm{H}=(H_x, 0, H_z)=H(\sin \theta_{H}, 0, \cos \theta_H)$ for $0\leq \theta_H \leq \pi/2$.

\subsection{Ferroaxial moment}
\label{sec: Ferroaxial moment}

\begin{table*}[htb!]
\caption{
The relationship between the irreducible representation (irrep.) and multipole activated in spinless and spinful spaces under the point groups $D_{\rm 4h}$ and $C_{\rm 4h}$. 
The sign of superscript in the irreducible representation denotes the time-reversal parity ($+$: even and $-$: odd). 
The multipole degrees of freedom belonging to the A$_{2g}^+$ representation correspond to the ferroaxial moment. 
}
\label{tab: irrep}
\centering
\renewcommand{\arraystretch}{1.2}
 \begin{tabular}{ccccccc}
 \hline 
 \ \ \ \ irrep. $(D_{\rm 4h})$ \ \ \ \ & $(C_{\rm 4h})$ & \ \  spinless \ \  &  multipole  & \ \ spinful \ \ & multipole  \\ \hline \hline
A$_{1g}^+$ & A$_{g}^+$ & 4 & $Q_0, Q_u, Q_4, Q_{4u}$  &  5 & $Q_{0}, 2Q_u, Q_4, Q_{4u}$\\ 
A$_{2g}^+$ & A$_{g}^+$ & 1 & $Q_{4z}^\alpha$  & 3 & $Q_{4z}^\alpha, G_z, G_z^\alpha$ \\ 
B$_{1g}^+$ & B$_{g}^+$ &  2 & $Q_{v}, Q_{4v}$ & 4 & $2Q_{v}, Q_{4v}, G_{xyz}$\\ 
B$_{2g}^+$ & B$_{g}^+$ & 2 &  $Q_{xy}, Q_{4z}^\beta$  & 4 & $2Q_{xy}, Q_{4z}^\beta,G_z^\beta$\\ 
E$_{g}^+$ & E$_{g}^+$ &  3 & $Q_{yz},Q_{4x}^\alpha, Q_{4x}^\beta$  & 7 & $2Q_{yz},Q_{4x}^\alpha, Q_{4x}^\beta, G_x, G_{3a},G_{3u}$\\ 
 &  &   & $Q_{zx}, Q_{4y}^{\alpha}, Q_{4y}^\beta$  &  & $2Q_{zx}, Q_{4y}^{\alpha}, Q_{4y}^\beta, G_y,G_{3b},G_{3v}$\\ 
 \hline
A$_{1g}^-$ & A$_{g}^-$ & 0 & --  & 4 & $M_{5u}, T_u, T_4, T_{4u}$\\ 
A$_{2g}^-$ & A$_{g}^-$ & 2 & $M_z, M_z^\alpha$ & 7 & $2M_z, 2M_z^\alpha, M^{\alpha1 }_{5z}, M^{\alpha 2}_{5z}, T_{4z}^\alpha$\\ 
B$_{1g}^-$ & B$_{g}^-$ & 1 &$M_{xyz}$  & 5 & $2M_{xyz}, M_{5v}, T_{v}, T_{4v}$\\ 
B$_{2g}^-$ & B$_{g}^-$ & 1 & $M_z^\beta$ & 5 & $2M_z^\beta, M_{5z}^{\beta}, T_{xy}, T_{4z}^\beta$\\ 
E$_{g}^-$ & E$_{g}^-$ &  3 & $M_x,M_x^\alpha,M_x^\beta$  & 12 & $2M_x,2M_{3a},2M_{3u}, M^{\alpha1}_{5x}, M^{\alpha2}_{5x}, M^{\beta}_{5x}, T_{yz}, T_{4x}^\alpha, T_{4x}^\beta$\\ 
 & &  & $M_y,M_y^{\alpha},M_y^\beta$  & & $2M_y,2M_{3b},2M_{3v}, M^{\alpha 1}_{5y}, M^{\alpha2}_{5y}, M^{\beta}_{5y}, T_{zx},T_{4y}^\alpha, T_{4y}^\beta$ \\ 
 \hline
\hline 
\end{tabular}
\end{table*}

In the model in Eq.~(\ref{eq: Ham}), there are $100=10\times 10$ independent electronic degrees of freedom in the Hilbert space, which are composed of five orbital and two spin degrees of freedom ($5 \times 2=10$).  
In spinless space, there are $25=5\times 5$ electronic degrees of freedom. 
As we consider $d$ orbital with the orbital angular momentum $l=2$, the spinless Hilbert space is spanned by the rank 0--4 multipoles~\cite{hayami2018microscopic, Hayami_PhysRevB.98.165110, Yatsushiro_PhysRevB.104.054412, kusunose2022generalization}: the electric monopole $(Q_0)$, magnetic dipole $(M_x, M_y, M_z)$, electric quadrupole $(Q_u, Q_{v}, Q_{yz}, Q_{zx}, Q_{xy})$, magnetic octupole $(M_{xyz}, M_{3a}, M_{3b}, M_z^\alpha, M_{3u}, M_{3v}, M_z^\beta)$, and electric hexadecapole $(Q_{4}, Q_{4u}, Q_{4v}, Q^{\alpha}_{4x}, Q^\alpha_{4y}, Q^\alpha_{4z}, Q^\beta_{4x}, Q^\beta_{4y}, Q^\beta_{4z})$, i.e., $1+3+5+7+9=25$.
Among them, only $Q^\alpha_{4z}$ belongs to the ${\rm A}^+_{2g}$ representation of $D_{\rm 4h}$ point group ($+$ in the superscript means the time-reversal even), which is related to the ferroaxial moment. 
The matrix of $\mathcal{H}_V$ in Eq.~(\ref{eq: HamV}) corresponds to $Q^\alpha_{4z}$. 
The other 75 electronic degrees of freedom activated in spinful Hilbert space are as follows: one electric monopole, two magnetic dipoles, two electric quadrupoles, two magnetic octupoles,   one electric hexadecapole, one magnetic dotriacontapole ($M_{5u}, M_{5v}, M^{\alpha1}_{5x}, M^{\alpha1}_{5y}, M^{\alpha1}_{5z}, M^{\alpha2}_{5x}, M^{\alpha2}_{5y}, M^{\alpha2}_{5z}, M^{\beta}_{5x}, M^{\beta}_{5y}, M^{\beta}_{5z}$), one electric toroidal dipole $(G_x, G_y, G_z)$, one magnetic toroidal quadrupole $(T_u, T_{v}, T_{yz}, T_{zx}, T_{xy})$, one electric toroidal octupole $(G_{xyz}, G_{3a}, G_{3b}, G_z^\alpha, G_{3u}, G_{3v}, G_z^\beta)$, and one magnetic toroidal hexadecapole $(T_{4}, T_{4u}, T_{4v}, T^{\alpha}_{4x}, T^\alpha_{4y}, T^\alpha_{4z}, T^\beta_{4x}, T^\beta_{4y}, T^\beta_{4z})$, i.e., $1+2\times 3 + 2\times 5 +2 \times 7+9+11+3+5+7+9=75$. 
Among them, there are three multipole degrees of freedom belonging to the ${\rm A}^+_{2g}$ representation of $D_{\rm 4h}$; $Q^\alpha_{4z}$, $G_z$, and $G^\alpha_z$.
The relationship between the irreducible representation under the point groups $D_{\rm 4h}$ and $C_{\rm 4h}$ and multipoles activated in spinless and spinful spaces are summarized in Table~\ref{tab: irrep}.  

The matrix elements of four multipoles belonging to the ${\rm A}^+_{2g}$ representation are given by 
\begin{widetext}
\begin{align}
& Q^{\alpha}_{4z} =
\begin{pmatrix}
0 & 0 & 0 & 0 & 0 \\
0 & 0 & 0 & 0 & 1 \\
0 & 0 & 0 & 0 & 0 \\
0 & 0 & 0 & 0 & 0 \\
0 & 1 & 0 & 0 & 0
\end{pmatrix}  \sigma_{0}, \\
&G_{z}=\frac{1}{2}
\begin{pmatrix}
  0 & 0 & \sqrt{3}i & 0 & 0 \\
  0 & 0 & i & 0 & 0 \\
  -\sqrt{3}i & -i & 0 & 0 & 0 \\
  0 & 0 & 0 & 0 & i \\
  0 & 0 & 0 & -i & 0
\end{pmatrix}  \sigma_{y}-\frac{1}{2}
\begin{pmatrix}
  0 & 0 & 0 & -\sqrt{3}i & 0 \\
  0 & 0 & 0 & i & 0 \\
  0 & 0 & 0 & 0 & -i \\
  \sqrt{3}i & -i & 0 & 0 & 0 \\
  0 & 0 & i & 0 & 0
\end{pmatrix} \sigma_x, \\
&G^\alpha_{z}=
\begin{pmatrix}
0 & 0 & 2i & 0 & 0 \\
0 & 0 & -\sqrt{3}i & 0 & 0 \\
-2i & \sqrt{3}i & 0 & 0 & 0 \\
0 & 0 & 0 & 0 & -\sqrt{3}i \\
0 & 0 & 0 & \sqrt{3}i & 0
\end{pmatrix}  \sigma_{y}-
\begin{pmatrix}
0 & 0 & 0 & -2i & 0 \\
0 & 0 & 0 & -\sqrt{3}i & 0 \\
0 & 0 & 0 & 0 & \sqrt{3}i \\
2i & \sqrt{3}i & 0 & 0 & 0 \\
0 & 0 & -\sqrt{3}i & 0 & 0
\end{pmatrix}  \sigma_x, \\
&Q'^\alpha_{4z}=
\begin{pmatrix}
0 & 0 & 0 & 0 & 0 \\
0 & 0 & -i & 0 & 0 \\
0 & i & 0 & 0 & 0 \\
0 & 0 & 0 & 0 & i \\
0 & 0 & 0 & -i & 0
\end{pmatrix}  \sigma_{y}+
\begin{pmatrix}
  0 & 0 & 0 & 0 & 0 \\
  0 & 0 & 0 & i & 0 \\
  0 & 0 & 0 & 0 & i \\
  0 & -i & 0 & 0 & 0 \\
  0 & 0 & -i & 0 & 0
\end{pmatrix}  \sigma_x,
\end{align}
\end{widetext}
for the basis $(d_u, d_{v}, d_{yz}, d_{zx}, d_{xy})$; $\bm{\sigma}=(\sigma_x, \sigma_y, \sigma_z)$ is the vector of the Pauli matrices and $\sigma_0$ is the $2\times 2$ unit matrix in spin space~\cite{Hayami_doi:10.7566/JPSJ.91.113702,Kusunose_PhysRevB.107.195118}. 
To distinguish the electric hexadecapole in spinless and spinful bases, we denote the latter as $Q'^\alpha_{4z}$. 

By numerically calculating the expectation values of $(Q^\alpha_{4z},Q'^\alpha_{4z}, G_z, G_z^\alpha)$ for the Hamiltonian in Eq.~(\ref{eq: Ham}) at zero field, one finds that $\langle Q^\alpha_{4z} \rangle $, $\langle Q'^\alpha_{4z} \rangle$, and $\langle G_z^\alpha \rangle$ become nonzero but $\langle G_z \rangle $ vanishes ($\langle \cdots \rangle$ stands for the statistical average)~\cite{Hayami_doi:10.7566/JPSJ.91.113702}. 
In other words, the dipole component of the electric toroidal multipole is not activated in the model Hamiltonian in Eq.~(\ref{eq: Ham}) even when it belongs to the totally symmetric irreducible representation for $V \neq 0$. 

\begin{figure}[t!]
\begin{center}
\includegraphics[width=1.0 \hsize]{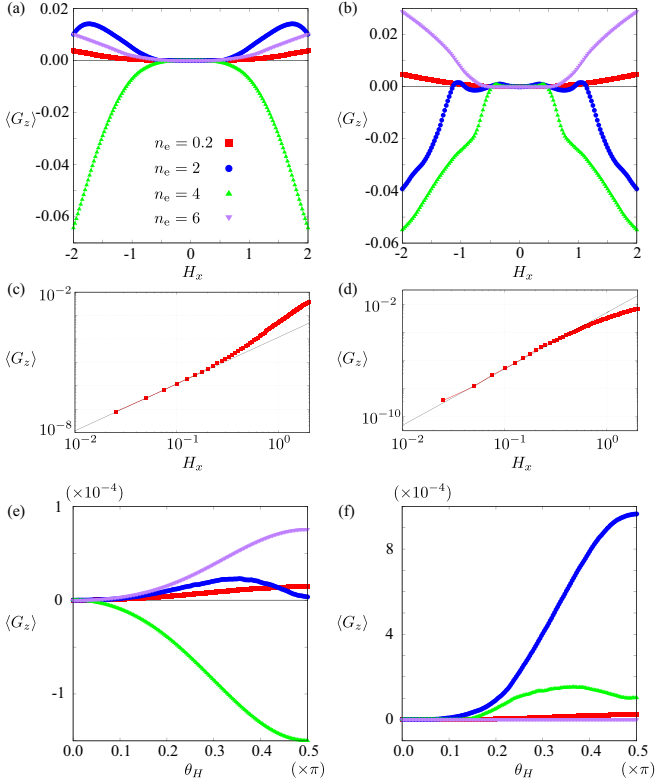}
\caption{
\label{Fig: MF_Gz}
(a,b) $H_x$ dependence of the expectation values of $\langle G_z \rangle $ for several electron fillings $n_{\rm e}$ at (a) $\lambda=2$ and (b) $\lambda=0$. 
(c, d) The double logarithmic plot for $H_x>0$ corresponding to (a) and (b) at $n_{\rm e}= 0.2$. 
The black lines are lines proportional to (c) $H_x^2$ and (d) $H_x^4$. 
(e,f) $\theta_H$ dependence of $\langle G_z \rangle $ at (e) $\lambda=2$ and (f) $\lambda=0$ for $H=0.3$ and several $n_{\rm e}$. 
}
\end{center}
\end{figure}

Meanwhile, we find that $\langle G_z \rangle $ becomes nonzero by introducing an external magnetic field in Eq.~(\ref{eq: HamZeeman}). 
Figure~\ref{Fig: MF_Gz}(a) shows the $H_x$ dependence of $\langle G_z \rangle $ for electron fillings $n_{\rm e}= 0.2$, $2$, $4$, and $6$; $n_{\rm e}=10$ means the full filling. 
We take the summation of the momentum $\bm{k}$ over $360^3$ grid points in the first Brillouin zone. 
The results at $n_{\rm e}= 0.2$ and $2$ are metallic and those at $n_{\rm e}= 4$ and $6$ are insulating. 
The data show that $\langle G_z \rangle $ becomes nonzero irrespective of metals and insulators for $H_x \neq 0$ at all $n_{\rm e}$ and its $H_x$ dependence is symmetric. 
Similarly, such behavior is also found in the case of $\lambda=0$, as shown in Fig.~\ref{Fig: MF_Gz}(b), which indicates that the spin--orbit coupling is not important in driving the field-induced electric toroidal dipole. 
Meanwhile, the $H_x$ dependence of $\langle G_z \rangle $ is different in each case; $\langle G_z \rangle $ is proportional to $H_x^2$ ($H_x^4$) for $\lambda \neq 0$ ($\lambda =0$), as shown in Fig.~\ref{Fig: MF_Gz}(c) [Fig.~\ref{Fig: MF_Gz}(d)]. 
One can analytically obtain such a difference by performing essential-model-parameter calculations in Appendix~\ref{sec: Essential model parameters}~\cite{Oiwa_doi:10.7566/JPSJ.91.014701}.

We also show the $\theta_H$ dependence of $\langle G_z \rangle $ for $H=0.3$ at $\lambda=2$ in Fig.~\ref{Fig: MF_Gz}(e) and at $\lambda=0$ in Fig.~\ref{Fig: MF_Gz}(f). 
The result indicates that $\langle G_z \rangle $ becomes nonzero except for $\theta_{H}=0$, i.e., the $z$ direction. 
Moreover, one finds that $\langle G_z \rangle $ tends to be larger as $\theta_{H}$ approaches $\pi/2$. 
Thus, the atomic-scale electric hexadecapole $\langle Q^{\alpha}_{4z} \rangle$ that arises from $V$ in Eq.~(\ref{eq: HamV}) is related to the electric toroidal dipole via the in-plane component of the applied magnetic field. 

The above result that $\langle G_z \rangle $ is symmetric against the magnetic field is understood from the time-reversal symmetry. 
Since the time-reversal parity of $G_z$ is $+1$, it can couple to the even order of the magnetic field, whose time-reversal parity is given as $-1$. 
Meanwhile, when the time-reversal symmetry of the system is broken so as to accommodate the time-reversal odd and spatial-inversion even multipoles, such as the magnetic toroidal monopole and quadrupole, $\langle G_z \rangle $ behaves antisymmetric against the magnetic field~\cite{hayami2023time}.

\section{Unconventional Hall effect}
\label{sec: Planar Hall effect}

In this section, we discuss the unconventional Hall effect under the ferroaxial ordering. 
We first discuss the general symmetry condition of the Hall effect in Sec.~\ref{sec: Tensor}. 
Next, we examine the numerical results by comparing the conventional Hall effect without the ferroaxial moment in Sec.~\ref{sec: Numerical results}. 

\subsection{Hall conductivity Tensor}
\label{sec: Tensor}

The Hall conductivity tensor $\sigma^{\rm H}_{\mu\nu;\eta}$ is defined by 
\begin{align}
J_{\mu}=\sum_{\nu \eta}\sigma^{\rm H}_{\mu\nu;\eta} E_\nu H_\eta, 
\end{align}
where $J_{\mu}$, $E_{\nu}$, and $H_\eta$ are electric current, electric field, and magnetic field for $\mu, \nu, \eta=(x,y,z)$, respectively. 
Since $\sigma^{\rm H}_{\mu\nu;\eta}$ corresponds to the antisymmetric part of the linear conductivity tensor proportional to $H$, it satisfies the relation of $\sigma^{\rm H}_{\mu\nu;\eta}=-\sigma^{\rm H}_{\nu\mu;\eta}$ from the Onsager reciprocal relations. 
Then, there are nine independent tensor components in $\sigma^{\rm H}_{\mu\nu;\eta}$, which are represented by 
\begin{align}
\label{eq: Hall}
\sigma^{\rm H}&=
\begin{pmatrix}
\sigma^{\rm H}_{yz;x}
 & \sigma^{\rm H}_{yz;y}
 & \sigma^{\rm H}_{yz;z} \\
\sigma^{\rm H}_{zx;x}
 &\sigma^{\rm H}_{zx;y}
 &\sigma^{\rm H}_{zx;z} \\
\sigma^{\rm H}_{xy;x}
 & \sigma^{\rm H}_{xy;y}
 &\sigma^{\rm H}_{xy;z} \\
\end{pmatrix} \nonumber \\
&=
\begin{pmatrix}
Q_0-Q_u+Q_{v} & Q_{xy}+G_z & Q_{zx} - G_y \\
Q_{xy} - G_z & Q_0 - Q_u-Q_{v} & Q_{yz}+G_x \\
Q_{zx} +G_y & Q_{yz}-G_x & Q_0+2Q_u \\
\end{pmatrix}.
\end{align}
Here, the second line of Eq.~(\ref{eq: Hall}) describes the component of $\sigma^{\rm H}$ in terms of multipoles. 
Since the antisymmetric $\sigma^{\rm H}_{\mu\nu;\eta}$ is the second-rank polar tensor, the relevant multipoles are the rank 0--2 time-reversal even multipoles, i.e., electric monopole $Q_0$, electric toroidal dipole $(G_x, G_y, G_z)$, and electric quadrupole $(Q_u, Q_{v}, Q_{yz}, Q_{zx}, Q_{xy})$~\cite{Hayami_PhysRevB.98.165110}. 
Once any of the multipoles belong to the totally symmetric irreducible representation under the point group we focus on, the corresponding tensor component can be nonzero.

In the case of the point group $D_{\rm 4h}$, $Q_0$ and $Q_u$ belong to the totally symmetric irreducible representation. 
Then, the nonzero tensor component is represented by 
\begin{align}
\label{eq:rank12}
\sigma^{\rm H}(D_{\rm 4h}) 
&=
\begin{pmatrix}
Q_0-Q_u & 0 & 0 \\
0 & Q_0 - Q_u & 0 \\
0 & 0 & Q_0+2Q_u \\
\end{pmatrix}. 
\end{align}
The induced component corresponds to the conventional Hall effect, where the transverse current occurs perpendicular to both the input electric field and the magnetic field. 

Meanwhile, when the symmetry is lowered from $D_{\rm 4h}$ to $C_{\rm 4h}$ so that the ferroaxial moment is induced, the component corresponding to the electric toroidal dipole $G_z$ can be nonzero. 
The additional tensor component is represented by 
\begin{align}
\label{eq:rank12_FAO}
\sigma^{\rm H}{\rm (FAO)}&=
\begin{pmatrix}
0 & G_z & 0 \\
- G_z & 0 & 0 \\
0 & 0 & 0 \\
\end{pmatrix}.
\end{align}
Thus, $\sigma^{\rm H}_{zx;x}=-\sigma^{\rm H}_{yz;y}$ is expected under the ferroaxial ordering. 
Both $\sigma^{\rm H}_{zx;x}$ and $\sigma^{\rm H}_{yz;y}$ mean the transverse conductivity under the magnetic field perpendicular to the ferroaxial moment; all $J_\mu$, $E_\nu$, and $H_\eta$ lie in the same plane, which corresponds to the unconventional Hall effect~\cite{Tan_PhysRevB.103.214438} rather than planar Hall effect~\cite{PhysRevLett.90.107201, PhysRevB.96.041110, PhysRevLett.119.176804, PhysRevResearch.3.L012006}. 

The emergence of the unconventional Hall effect under the ferroaxial ordering is intuitively understood from the nature of the ferroaxial moment as a nanometric rotator that causes the transverse responses between the conjugate physical quantities~\cite{cheong2021permutable, Hayami_doi:10.7566/JPSJ.91.113702}. 
We consider the case of $\sigma^{\rm H}_{zx;x}$ as an example. 
When the magnetic field is applied in the $x$ direction, the system exhibits the magnetization along the perpendicular $y$ direction in addition to the parallel $x$ direction~\cite{inda2023nonlinear}.  
Such an induced magnetization along $y$ axis can contribute to the ordinary Hall effect in the $zx$ plane. 
In the end, one can find the emergence of the unconventional Hall effect $\sigma^{\rm H}_{zx;x}$ under the ferroaxial ordering. 
In addition, one finds that $\sigma^{\rm H}_{zx;x}$ is an odd function of the magnetic field owing to the time-reversal parity of the ferroaxial ordering.
In the following, we focus on the behavior of $\sigma^{\rm H}_{zx;x}$ induced by the ferroaxial ordering. 

\subsection{Numerical results}
\label{sec: Numerical results}

\begin{figure}[htb!]
\begin{center}
\includegraphics[width=0.8 \hsize]{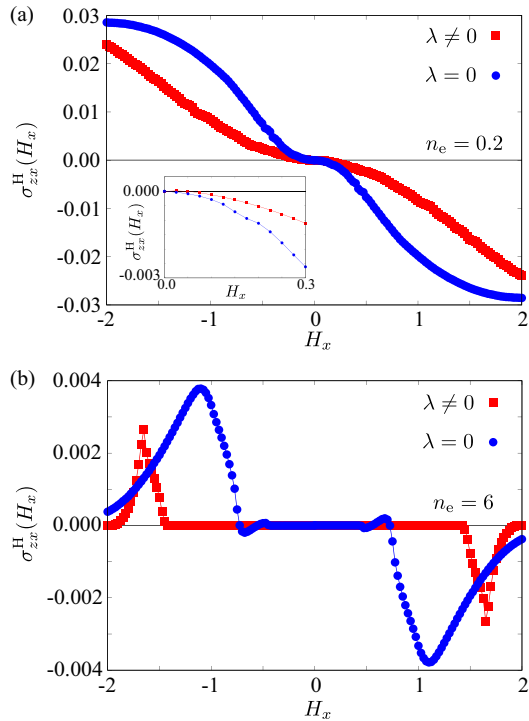}
\caption{
\label{Fig: PHall_Hdep}
$H_x$ dependence of the unconventional Hall conductivity $\sigma^{\rm H}_{zx}(H_{x})$ at $\lambda=0$ and 2 for (a) $n_{\rm e}=0.2$ and (b) $n_{\rm e}= 6$. 
The inset of (a) represents $\sigma^{\rm H}_{zx}(H_{x})$ for small positive $H_x$.
}
\end{center}
\end{figure}

We compute $\sigma^{\rm H}_{\mu \nu}(H_{\eta})$ by using the Kubo formula within the linear response theory~\cite{kubo_doi:10.1143/JPSJ.12.570} as 
\begin{align}
\label{eq: Kubo}
\sigma^{\rm H}_{\mu \nu}(H_{\eta}) =\frac{e^2}{\hbar}\frac{1}{i \bar{V}} \sum_{m,n,\bm{k}} 
\frac{f(\epsilon_{n \bm{k}})-f(\epsilon_{m \bm{k}})}{\epsilon_{n \bm{k}}-\epsilon_{m \bm{k}}} 
\frac{J_{\mu\bm{k}}^{nm} J_{\nu\bm{k}}^{mn}
}{\epsilon_{n \bm{k}}-\epsilon_{m \bm{k}}+{i \delta}}, 
\end{align}
where $e$ is the elementary charge, $\hbar=h/(2\pi)$ is the reduced Planck constant, $\bar{V}$ is the system volume, $\delta$ is the broadening factor, $f(\epsilon)$ is the Fermi distribution function, and $J_{\mu\bm{k}}^{nm}=\langle n\bm{k} | J_{\mu} | m \bm{k} \rangle$ is the matrix element of the current operator in the direction $\mu=(x,y,z)$, $J_{\mu}$. 
$\epsilon_{m \bm{k}}$ and $|m \bm{k} \rangle$ are the $m$-th eigenvalue and eigenstate of $\mathcal{H}$, respectively; we include the effect of an external magnetic field as the Zeeman coupling in $\mathcal{H}$ ($\eta$ represents the magnetic-field direction), and hence, the higher-order contribution of $\bm{H}$ is included in the conductivity tensor. 
We set $e^2/h=1$, $\delta = 0.01$ and temperature $T=0.001$. 
The summation of $\bm{k}$ is taken over the $360^3$ grid points in the Brillouin zone. 

Figure~\ref{Fig: PHall_Hdep}(a) shows the $H_x$ dependence of $\sigma^{\rm H}_{zx}(H)$ at $\lambda=0$ and 2 for $n_{\rm e}=0.2$. 
As expected from the symmetry consideration in Sec.~\ref{sec: Tensor}, $\sigma^{\rm H}_{zx}(H)$ becomes nonzero and is an odd function of $H_x$, as the conventional Hall effect. 
We also confirm the relation of $\sigma^{\rm H}_{zx}(H_{x})=-\sigma^{\rm H}_{yz}(H_{y})$, although $\sigma^{\rm H}_{zx}(H_{x})$ is not linear to $H_x$ for small $H_x$, as shown in the inset of Fig.~\ref{Fig: PHall_Hdep}(a); this might be ascribed to the higher-order effect in terms of $H_x$ included in Eq.~(\ref{eq: Kubo}). 
Furthermore, one finds that the spin--orbit coupling $\lambda$ does not play 
an intrinsic role in inducing $\sigma^{\rm H}_{zx}(H_{x})$, as shown in the comparison of the results at $\lambda=0$ and 2. 
Since a larger Hall response is obtained for larger $|H_x|$, a larger magnetic field is preferred in order to detect it in experiments.
Note that the interband process with $\epsilon_{m \bm{k}} \neq \epsilon_{n \bm{k}}$ in Eq.~(\ref{eq: Kubo}) contributes to $\sigma^{\rm H}_{zx}(H_{x})$. 

In order to analytically obtain the essential model parameters to cause $\sigma^{\rm H}_{zx}(H_{x})$, we perform an expansion method to $\sigma^{\rm H}_{zx}(H_{x})$~\cite{Oiwa_doi:10.7566/JPSJ.91.014701}. 
As a result, we find that $\sigma^{\rm H}_{zx}(H_{x})$ is proportional to $V H^{2m+1}_x g(t_{z2\sigma}, t_{z2\pi})$ ($m$ is integer and $g(t_{z2\sigma}, t_{z2\pi})$ is an appropriate function depending on $t_{z2\sigma}$ and $t_{z2\pi}$; $g(0,0)=0$, which is consistent with both the symmetry and the numerical results. 
From the above expression, we find that the next-nearest-neighbor hopping along the $z$ direction ($t_{z2\sigma}, t_{z2\pi}$) is important to obtain nonzero $\sigma^{\rm H}_{zx}(H_{x})$. 
The details of the expansion method are presented in Appendix~\ref{sec: Essential model parameters}. 

A similar behavior is found in the insulating case at $n_{\rm e}=6$, as shown in Fig.~\ref{Fig: PHall_Hdep}(b). 
Although $\sigma^{\rm H}_{zx}(H_{x})=0$ for $|H_x| \lesssim 1.4$ at $\lambda=2$ and for $|H_x| \lesssim 0.45$ at $\lambda=0$ in the insulating region, $H_{x}$-odd behavior becomes prominent once the band gap is closed by a larger magnetic field. 
In other words, it is difficult to observe the Hall response in the insulating case in experiments, since the large magnetic field comparable to the band gap is necessary; such an issue is evaded by considering the optical Hall conductivity by applying an ac electric field instead of a dc one.
We also present the result of the optical unconventional Hall conductivity, which can give a signal of ferroaxial ordering in the insulating case, in Appendix~\ref{sec: Optical conductivity}. 

\begin{figure}[htb!]
\begin{center}
\includegraphics[width=0.8 \hsize]{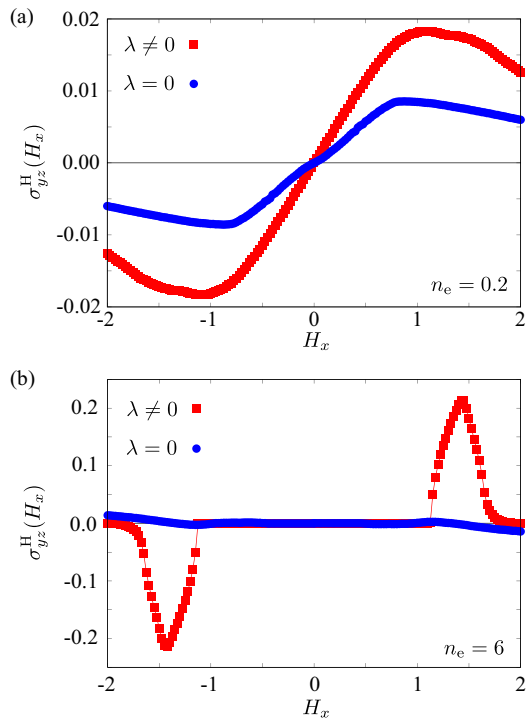}
\caption{
\label{Fig: NHall_Hdep}
$H_x$ dependence of the conventional Hall conductivity $\sigma^{\rm H}_{yz}(H_{x})$ at $\lambda=0$ and 2 for (a) $n_{\rm e}=0.2$ and (b) $n_{\rm e}= 6$. 
}
\end{center}
\end{figure}

We also calculate the ordinary nonplanar Hall conductivity $\sigma^{\rm H}_{yz}(H_{x})$ while the field direction is kept. 
Figures~\ref{Fig: NHall_Hdep}(a) and \ref{Fig: NHall_Hdep}(b) show the $H_x$ dependence of $\sigma^{\rm H}_{yz}(H_{x})$ at $n_{\rm e}=0.2$ and $n_{\rm e}=6$, respectively. 
As compared to the result of the unconventional Hall effect in Fig.~\ref{Fig: PHall_Hdep}, the behavior is similar to each other; $\sigma^{\rm H}_{yz}(H_{x})$ becomes nonzero for $H_x \neq 0$ and an odd function of $H_x$. 
On the other hand, we find the difference in their essential model parameters; the unconventional Hall conductivity $\sigma^{\rm H}_{zx}(H_{x})$ needs the CEF parameter $V$, while the conventional Hall conductivity $\sigma^{\rm H}_{yz}(H_{x})$ does not, as detailed in Appendix~\ref{sec: Essential model parameters}.

\section{Magnetoconductivity}
\label{sec: Magnetoconductivity}

We discuss the magnetoconductivity under the ferroaxial ordering. 
Similarly to the unconventional Hall effect in Sec.~\ref{sec: Planar Hall effect}, we show the symmetry condition and the relation to the multipoles in Sec.~\ref{sec: Tensor_2}. 
Then, we compute the magnetoconductivity tensor based on the Kubo formula in Sec.~\ref{sec: Numerical results_2}.

\subsection{Magnetoconductivity Tensor}
\label{sec: Tensor_2}

The magnetoconductivity tensor corresponding to the rank-4 polar tensor is defined by 
\begin{align}
J_{\mu}=\sum_{\nu \eta \gamma}\sigma^{\rm MC}_{\mu\nu;\eta\gamma} E_\nu H_\eta H_{\gamma}, 
\end{align}
where $\sigma^{\rm MC}_{\mu\nu;\eta\gamma}$ is the symmetric tensor as $\sigma^{\rm MC}_{\mu\nu;\eta\gamma}=\sigma^{\rm MC}_{\nu\mu;\eta\gamma}=\sigma^{\rm MC}_{\mu\nu;\gamma\eta}$. 
The independent 36 components in $\sigma^{\rm MC}_{\mu\nu;\eta\gamma}$ are represented by the $6\times 6$ matrix as follows: 
\begin{align}
\label{eq: MCtensor}
\sigma^{\rm MC}&=
\begin{pmatrix}
\sigma^{\rm MC}_{xx;xx}
 & \sigma^{\rm MC}_{xx;yy}
 & \sigma^{\rm MC}_{xx;zz}
 & \sigma^{\rm MC}_{xx;yz}
 &\sigma^{\rm MC}_{xx;zx}
 &\sigma^{\rm MC}_{xx;xy}
 \\
\sigma^{\rm MC}_{yy;xx}
 & \sigma^{\rm MC}_{yy;yy}
 & \sigma^{\rm MC}_{yy;zz}
 & \sigma^{\rm MC}_{yy;yz}
 & \sigma^{\rm MC}_{yy;zx}
 & \sigma^{\rm MC}_{yy;xy}
 \\
\sigma^{\rm MC}_{zz;xx}
 & \sigma^{\rm MC}_{zz;yy}
 & \sigma^{\rm MC}_{zz;zz}
 & \sigma^{\rm MC}_{zz;yz}
 & \sigma^{\rm MC}_{zz;zx}
 & \sigma^{\rm MC}_{zz;xy}  \\
 \sigma^{\rm MC}_{yz;xx} 
 &  \sigma^{\rm MC}_{yz;yy}
 & \sigma^{\rm MC}_{yz;zz}
 &
 \sigma^{\rm MC}_{yz;yz}
 &  \sigma^{\rm MC}_{yz;zx}
 &  \sigma^{\rm MC}_{yz;xy}\\
 \sigma^{\rm MC}_{zx;xx}
 &  \sigma^{\rm MC}_{zx;yy}
 &  \sigma^{\rm MC}_{zx;zz}
 &   \sigma^{\rm MC}_{zx;yz} 
 &  \sigma^{\rm MC}_{zx;zx}
 &  \sigma^{\rm MC}_{zx;xy} \\
  \sigma^{\rm MC}_{xy;xx}
 &   \sigma^{\rm MC}_{xy;yy} 
 &   \sigma^{\rm MC}_{xy;zz}
 &   \sigma^{\rm MC}_{xy;yz}
 &   \sigma^{\rm MC}_{xy;zx}
 &   \sigma^{\rm MC}_{xy;xy}
 \\
 \end{pmatrix} \nonumber \\
 &=\sigma^{\rm MC (M)}+\sigma^{\rm MC (D)}
 +\sigma^{\rm MC (Q)}
 +\sigma^{\rm MC (O)}
+ \sigma^{\rm MC (H)},
\end{align}
where $\sigma^{\rm MC (M)}$, $\sigma^{\rm MC (D)}$, $\sigma^{\rm MC (Q)}$, $\sigma^{\rm MC (O)}$,  and $\sigma^{\rm MC (H)}$ denote the monopole, dipole, quadrupole, octupole, and hexadecapole components in $\sigma^{\rm MC}$. 
The correspondence between the tensor components and multipoles is given by 
\begin{widetext}
\begin{align}
\label{eq: MCtensor_M}
\sigma^{\rm MC (M)}&=
\begin{pmatrix}
4Q_0+Q'_0 
 & -2Q_0+Q'_0
 & -2Q_0+Q'_0
 & 0 
 &0
 &0 
 \\
 -2Q_0+Q'_0
 & 4Q_0+Q'_0  
 & -2Q_0+Q'_0
 & 0
 & 0
 & 0 
 \\
-2Q_0+Q'_0
 & -2Q_0+Q'_0
 & 4Q_0+Q'_0 
 & 0
 & 0
 & 0  \\
 0 
 & 0 
 &0 
 &
 3 {Q}_{0}
 & 0 
 & 0\\
0
 & 0
 & 0
 &0 
 & 3 {Q}_{0}
 & 0 \\
 0 
 & 0 
 & 0 
 & 0 
 & 0 
 & 3 {Q}_{0} 
 \\
 \end{pmatrix}, \\
\label{eq: MCtensor_D}
\sigma^{\rm MC (D)}&=
\begin{pmatrix}
0 
 & 0 
 & 0
 & 0 
 &-2 {G}_{y} 
 & 2 {G}_{z}
 \\
 0
 & 0 
 & 0 
 &2 {G}_{x} 
 & 0 
 & -2 {G}_{z}
 \\
0
 & 0
 & 0 
 & -2 {G}_{x} 
 & 2 {G}_{y} 
 & 0 \\
 0 
 & -2 {G}_{x} 
 &2 {G}_{x} 
 & 0
 & -{G}_{z} 
 & {G}_{y}\\
2 {G}_{y}
 & 0 
 & -2 {G}_{y} 
 &{G}_{z} 
 & 0
 & -{G}_{x}\\
 -2 {G}_{z}
 & 2 {G}_{z} 
 & 0 
 & -{G}_{y} 
 & {G}_{x} 
 & 0\\
 \end{pmatrix}, \\
 \label{eq: MCtensor_Q}
\sigma^{\rm MC (Q)}&=
\begin{pmatrix}
\tilde{Q}_u+\tilde{Q}_v
 & -\tilde{Q}_u^{(+)}-\tilde{Q}_u^{(-)} + 2 \tilde{Q}_{v}^{(-)}
 & \tilde{Q}_u^{(+)}+\tilde{Q}_v^{(-)}
 &\tilde{Q}_{yz}^{(+)}
 &\tilde{Q}_{zx}^{\prime(+)} 
 & \tilde{Q}_{xy}^{\prime(+)} 
 \\
-\tilde{Q}_u^{(+)}-\tilde{Q}_u^{(-)}+2 \tilde{Q}_{v}^{(+)} 
 & \tilde{Q}_u-\tilde{Q}_v
 & \tilde{Q}_u^{(+)}-\tilde{Q}_v^{(-)}
 &\tilde{Q}_{yz}^{\prime(+)} 
 & \tilde{Q}_{zx}^{(+)}
 & \tilde{Q}_{xy}^{\prime(+)}
 \\
\tilde{Q}_u^{(-)}+\tilde{Q}_v^{(+)}
 & \tilde{Q}_u^{(-)}-\tilde{Q}_v^{(+)}
 & -2\tilde{Q}_u 
 &\tilde{Q}_{yz}^{\prime(+)} 
 & \tilde{Q}_{zx}^{\prime(+)} 
 & \tilde{Q}_{xy}^{(+)}  \\
 \tilde{Q}_{yz}^{(-)} 
 & \tilde{Q}_{yz}^{\prime(-)} 
 &\tilde{Q}_{yz}^{\prime(-)} 
 &3 {Q}_{u}-3 {Q}_{v}
 & 3 {Q}_{xy} 
 & 3 {Q}_{zx} \\
\tilde{Q}_{zx}^{\prime(-)}  
 & \tilde{Q}_{zx}^{(-)} 
 & \tilde{Q}_{zx}^{\prime(-)} 
 & 3 {Q}_{xy} 
 & 3 {Q}_{u}+3 {Q}_{v}
 & 3 {Q}_{yz} \\
\tilde{Q}_{xy}^{\prime(-)} 
 & \tilde{Q}_{xy}^{\prime(-)} 
 & \tilde{Q}_{xy}^{(-)} 
 & 3 {Q}_{zx}
 & 3 {Q}_{yz} 
 & -6 {Q}_{u}\\
 \end{pmatrix}, \\
 \label{eq: MCtensor_O}
\sigma^{\rm MC (O)}&=
\begin{pmatrix}
0
 & {G}_{xyz}
 & -{G}_{xyz}
 &-2 {G}_{x}^\beta
 &{G}_{y}^\alpha+{G}_{y}^\beta 
 & -{G}_{z}^\alpha+{G}_{z}^\beta
 \\
 -{G}_{xyz}
 & 0 
 & {G}_{xyz}
 &-{G}_{x}^\alpha+{G}_{x}^\beta
 & -2 {G}_{y}^\beta
 & {G}_{z}^\alpha+{G}_{z}^\beta
 \\
{G}_{xyz}
 & -{G}_{xyz}
 & 0 
 &{G}_{x}^\alpha+{G}_{x}^\beta
 & -{G}_{y}^\alpha+{G}_{y}^\beta 
 & -2 {G}_{z}^\beta \\
 2 {G}_{x}^\beta
 & {G}_{x}^\alpha-{G}_{x}^\beta 
 & - {G}_{x}^\alpha-{G}_{x}^\beta 
 & 0
 & -2 {G}_{z}^\alpha
 & 2 {G}_{y}^\alpha\\
-{G}_{y}^\alpha-{G}_{y}^\beta 
 & 2 {G}_{y}^\beta 
 & {G}_{y}^\alpha-{G}_{y}^\beta 
 &2 {G}_{z}^\alpha 
 & 0
 & -2 {G}_{x}^\alpha \\
{G}_{z}^\alpha-{G}_{z}^\beta 
 & -{G}_{z}^\alpha-{G}_{z}^\beta
 & 2 {G}_{z}^\beta 
 &  -2 {G}_{y}^\alpha
 & 2 {G}_{x}^\alpha
 & 0 \\
 \end{pmatrix}, \\
 \label{eq: MCtensor_H}
\sigma^{\rm MC (H)}&=
\begin{pmatrix}
2 {Q}_{4}-{Q}_{4u}+{Q}_{4v} 
 & -{Q}_{4}+2 {Q}_{4u} 
 &- Q_{4}-Q_{4u}- Q_{4v}
 &2 {Q}_{4x}^{\beta} 
 &-{Q}_{4y}^\alpha-{Q}_{4y}^{\beta} 
 & {Q}_{4z}^\alpha-{Q}_{4z}^{\beta}
 \\
 -{Q}_{4}+2 {Q}_{4u} 
 & 2 {Q}_{4}-{Q}_{4u}-{Q}_{4v} 
 & -Q_{4}-Q_{4u}+ Q_{4v}
 &{Q}_{4x}^\alpha-{Q}_{4x}^{\beta} 
 & 2 {Q}_{4y}^{\beta} 
 & -{Q}_{4z}^\alpha-{Q}_{4z}^{\beta} 
 \\
-Q_{4}-Q_{4u}- Q_{4v}
 & -Q_{4}-Q_{4u}+ Q_{4v}
 & 2 {Q}_{4}+2 {Q}_{4u} 
 &  -{Q}_{4x}^\alpha-{Q}_{4x}^{\beta} 
 & {Q}_{4y}^\alpha-{Q}_{4y}^{\beta} 
 & 2 {Q}_{4z}^{\beta}  \\
 2 {Q}_{4x}^{\beta} 
 & {Q}_{4x}^\alpha-{Q}_{4x}^{\beta} 
 &-{Q}_{4x}^\alpha-{Q}_{4x}^{\beta} 
 & - Q_{4}-Q_{4u}+ Q_{4v}
 & 2 {Q}_{4z}^{\beta} 
 & 2 {Q}_{4y}^{\beta}\\
-{Q}_{4y}^\alpha-{Q}_{4y}^{\beta} 
 & 2 {Q}_{4y}^{\beta} 
 & {Q}_{4y}^\alpha-{Q}_{4y}^{\beta} 
 &2 {Q}_{4z}^{\beta} 
 & -Q_{4}-Q_{4u}- Q_{4v}
 & 2 {Q}_{4x}^{\beta}\\
 {Q}_{4z}^\alpha-{Q}_{4z}^{\beta} 
 & -{Q}_{4z}^\alpha-{Q}_{4z}^{\beta} 
 & 2 {Q}_{4z}^{\beta} 
 &  2 {Q}_{4y}^{\beta} 
 & 2 {Q}_{4x}^{\beta} 
 & -{Q}_{4}+2 {Q}_{4u}\\
 \end{pmatrix},
\end{align}
\end{widetext}
where 
\begin{align}
(\tilde{Q}_u, \tilde{Q}_u^{(\pm)}) &\equiv (-4 {Q}_{u}-2 {Q}_{u}^{(+)},  -4 {Q}_{u}+{Q}_{u}^{(+)} \pm 3 {Q}_{u}^{(-)}), \notag\\
(\tilde{Q}_v, \tilde{Q}_v^{(\pm)}) &\equiv (4 {Q}_{v}+2 {Q}^{(+)}_{v}, -4 {Q}_{v}+{Q}^{(+)}_{v} \pm {Q}^{(-)}_{v}),  \notag\\
(\tilde{Q}_{\zeta}^{(\pm)}, \tilde{Q}_{\zeta}^{\prime(\pm)}) &\equiv (-4Q_{\zeta}+Q^{(+)}_{\zeta}\pm Q^{(-)}_{\zeta}, 2Q_{\zeta}+Q^{(+)}_{\zeta}\pm Q^{(-)}_{\zeta}),
\end{align}
for $\zeta=(yz, zx, xy)$. 
Thus, rank-0--4 multipoles can contribute to $\sigma^{\rm MC}$.

For the point group $D_{\rm 4h}$, the multipoles belonging to the totally symmetric irreducible representation up to rank 4 are $Q_0$, $Q_u$, $Q_4$, and $Q_{4u}$. 
Then, nonzero components of $\sigma^{\rm MC}$ under $D_{\rm 4h}$ is given by 
\begin{align}
\label{eq: MCtensor_para}
\sigma^{\rm MC}(D_{\rm 4h})&=
\begin{pmatrix}
Q_1
 & Q_3
 & Q_4
 & 0 
 &0
 &0 
 \\
Q_3
 & Q_1
 & Q_4
 & 0
 & 0
 & 0 
 \\
Q_5
 & Q_5
 & Q_2
 & 0
 & 0
 & 0  \\
 0 
 & 0 
 &0 
 &
 Q_6
 & 0 
 & 0\\
0
 & 0
 & 0
 &0 
 & Q_6
 & 0 \\
 0 
 & 0 
 & 0 
 & 0 
 & 0 
 & Q_7
 \\
 \end{pmatrix}, 
\end{align}
where 
\begin{align}
Q_1&=4Q_0+Q'_0  + \tilde{Q}_u +2 {Q}_{4}-{Q}_{4u}, \\
Q_2&=4Q_0+Q'_0 -2\tilde{Q}_u +2 {Q}_{4}+2 {Q}_{4u}, \\ 
Q_3&=-2Q_0+Q'_0-\tilde{Q}_u^{(+)}-\tilde{Q}_u^{(-)}-{Q}_{4}+2 {Q}_{4u}, \\ 
Q_4&=-2Q_0+Q'_0+\tilde{Q}_u^{(+)}- Q_{4}-Q_{4u}, \\
Q_5&=-2Q_0+Q'_0+\tilde{Q}_u^{(-)}-Q_{4}-Q_{4u}, \\ 
Q_6&=3 {Q}_{0}+3 {Q}_{u}- Q_{4}-Q_{4u}, \\ 
Q_7&=3 {Q}_{0} -6 {Q}_{u}-{Q}_{4}+2 {Q}_{4u}. 
\end{align}
There are seven independent components of $\sigma^{\rm MC}$. 

For the point group $C_{\rm 4h}$, the electric toroidal octupole $G^{\alpha}_{z}$ and electric hexadecapole $Q^{\alpha}_{4z}$ belong to the totally symmetric irreducible representation in addition to the electric toroidal dipole $G_{z}$. 
Then, the additional nonzero components of $\sigma^{\rm MC}$ are given by 
\begin{align}
\label{eq: MCtensor_FAO}
\sigma^{\rm MC}{\rm (FAO)}&=
\begin{pmatrix}
0 
 & 0 
 & 0
 & 0 
 &0
 & G_2
 \\
 0
 & 0 
 & 0 
 &0
 & 0 
 & -G_2
 \\
0
 & 0
 & 0 
 & 0
 & 0
 & 0 \\
 0 
 & 0
 &0
 & 0
 & -G_3
 & 0\\
0
 & 0 
 & 0
 &G_3
 & 0
 & 0\\
 G_1
 & -G_1
 & 0 
 & 0
 & 0
 & 0\\
 \end{pmatrix}, 
\end{align}
where $G_1=-(2 {G}_{z}-{G}_{z}^\alpha-{Q}_{4z}^\alpha)$, $G_2=2 {G}_{z}-{G}_{z}^\alpha+{Q}_{4z}^\alpha$, and $G_3={G}_{z} +2 {G}_{z}^\alpha $.
Thus, $\sigma^{\rm MC}{\rm (FAO)}$ has three independent components: $\sigma^{\rm MC}_{xy;xx}=-\sigma^{\rm MC}_{xy;yy}$, $\sigma^{\rm MC}_{xx;xy}=-\sigma^{\rm MC}_{yy;xy}$, and $\sigma^{\rm MC}_{zx;yz}=-\sigma^{\rm MC}_{yz;zx}$. 
We focus on $\sigma^{\rm MC}_{xy;xx}$ in the following calculations, since other components, $\sigma^{\rm MC}_{xx;xy}$ and $\sigma^{\rm MC}_{zx;yz}$, also show a similar behavior to $\sigma^{\rm MC}_{xy;xx}$.

\subsection{Numerical results}
\label{sec: Numerical results_2}

\begin{figure}[htb!]
\begin{center}
\includegraphics[width=0.8 \hsize]{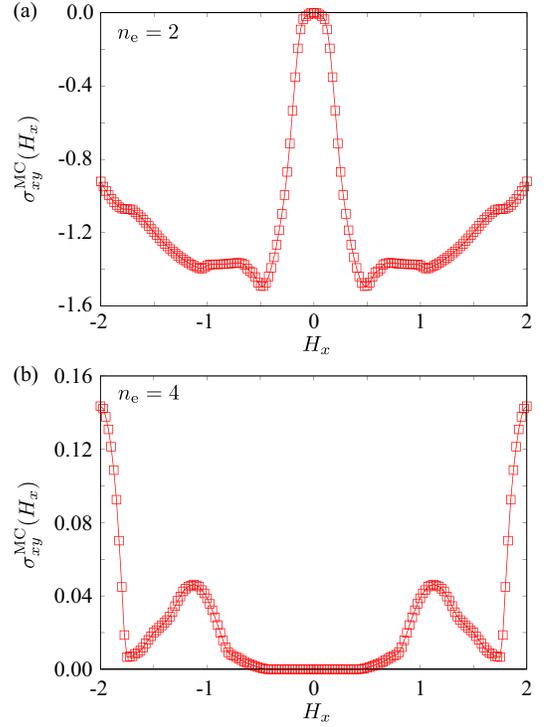}
\caption{
\label{Fig: MR1_Hxdep}
$H_x$ dependence of $\sigma^{\rm MC}_{xyxx}$ at $\lambda=2$ for (a) $n_{\rm e}=2$ and (b) $n_{\rm e}= 4$. 
}
\end{center}
\end{figure}

We calculate $\sigma^{\rm MC}$ by using the Kubo formula in Eq.~(\ref{eq: Kubo}), where the effect of the magnetic field is included in the Hamiltonian. 
In contrast to the unconventional Hall effect, the intraband process proportional to $1/\delta$ contributes to the magnetoconductivity tensor.  
Figure~\ref{Fig: MR1_Hxdep}(a) shows the $H_x$ dependence of $\sigma^{\rm MC}_{xy}(H_x)$ in the metallic ferroaxial ordered state at $n_{\rm e}=2$ for $\lambda=0$ and $2$. 
Similarly to the unconventional Hall effect in Sec.~\ref{sec: Numerical results}, $\sigma^{\rm MC}_{xy}(H_x)$ becomes nonzero for $H_x \neq 0$ irrespective of $\lambda$. 
On the other hand, it is the even function of $H_{x}$ owing to the opposite time-reversal parity of $\sigma^{\rm H}$. 
Indeed, the essential model parameters of $\sigma^{\rm MC}_{xy}(H_x)$ are obtained in the form of $V H^{2m}_x$, as shown in Appendix~\ref{sec: Essential model parameters}. 
These features are common to the insulating ferroaxial ordered state, as shown in Fig.~\ref{Fig: MR1_Hxdep}(b).

\section{Summary}
\label{sec: Summary}

To summarize, we have investigated metallic ferroaxial ordering, which is characterized by mirror symmetry breaking without the breaking of spatial inversion and time-reversal symmetries, by focusing on its transport property in the external magnetic field. 
We analyzed a fundamental five $d$-orbital model to include the electric toroidal dipole, electric toroidal octupole, and electric hexadecapole degrees of freedom corresponding to the ferroaxial moment under the tetragonal lattice structure. 
As a result, we find three characteristic features under the magnetic field. 
First, the electric toroidal dipole is induced by the in-plane magnetic field. 
Second, the unconventional Hall effect is induced as an odd function of the applied magnetic field. 
Third, the magnetoconductivity as an even function of the applied magnetic field occurs. 
In all the cases, we derive the essential model parameters out of the hoppings, spin--orbit coupling, and crystalline electric field. 
Our results indicate that the metallic ferroaxial ordered state becomes the source of unconventional magnetotransport phenomena in which the spin-orbit coupling is not necessary; it is noted that similar transport phenomena can be also expected in the insulating ferroaxial ordered state by applying an ac electric field. 
The present transport property can be detected in various materials without mirror symmetry parallel to the ferroaxial moment, such as Co$_3$Nb$_2$O$_8$~\cite{Johnson_PhysRevLett.107.137205}, CaMn$_7$O$_{12}$~\cite{Johnson_PhysRevLett.108.067201}, RbFe(MoO$_4$)$_2$~\cite{jin2020observation,Hayashida_PhysRevMaterials.5.124409}, NiTiO$_3$~\cite{hayashida2020visualization, Hayashida_PhysRevMaterials.5.124409, yokota2022three, Guo_PhysRevB.107.L180102}, Ca$_5$Ir$_3$O$_{12}$~\cite{Hasegawa_doi:10.7566/JPSJ.89.054602, hanate2021first, hayami2023cluster}, BaCoSiO$_4$~\cite{Xu_PhysRevB.105.184407}, K$_2$Zr(PO$_4$)$_2$~\cite{yamagishi2023ferroaxial}, Na$_2$Hf(BO$_3$)$_2$~\cite{nagai2023chemicalSwitching}, and Na-superionic conductors~\cite{nagai2023chemical}.
By using tight-binding model parameters obtained from density functional theory  (DFT) calculations, one can quantitatively evaluate the transport tensors in these materials.
Since such model parameters were obtained for Ca$_5$Ir$_3$O$_{12}$~\cite{Charlebois_PhysRevB.104.075153}, it is intriguing to examine the transport behavior in the ferroaxial system, which is left for future study.

\begin{acknowledgments}
This research was supported by JSPS KAKENHI Grants Numbers JP21H01031, JP21H01037, JP22H04468, JP22H00101, JP22H01183, JP23K03288, JP23H04869, JP23H00091 and by JST PRESTO (JPMJPR20L8).
Parts of the numerical calculations were performed in the supercomputing systems in ISSP, the University of Tokyo.
\end{acknowledgments}

\appendix

\section{Essential model parameters}
\label{sec: Essential model parameters}

\subsection{Hamiltonian in terms of electronic multipoles}
\label{sec: Hamiltonian in terms of electronic multipoles}

In this section, we present the model Hamiltonian expressed as the linear combination of the symmetry-adapted multipole basis (SAMB)~\cite{Oiwa_PhysRevLett.129.116401, Kusunose_PhysRevB.107.195118}, which enables us to clarify the hidden multipole degrees of freedom in the present system.
First, we introduce the atomic multipole basis defined in the spinful $d$ orbitals at a single site, and the momentum multipole basis defined as the function of the wave vector $\bm{k}$~\cite{Hayami_PhysRevB.98.165110, kusunose2020complete, Hayami_PhysRevB.102.144441}.
Then, we show that the SAMB is decomposed as the linear combination of the products of these two multipole bases.

\subsubsection{Atomic multipole basis \label{sec:am}}

\begin{table*}[hbt!]
  \caption{
  Operator expressions of the atomic multipole basis in the point group $D_{\rm 4h}$.
  Only the multipoles appearing in the Hamiltonian and some relevant ones are summarized.
  E, M, ET, and MT stand for electric, magnetic, electric toroidal, and magnetic toroidal, respectively.
  The superscript (a) denotes the atomic multipole.
  $\bm{l}$ and $\bm{\sigma}/2$ represent the dimensionless orbital and spin angular-momentum operators.
  The upper and lower parts separated by the double line represent the spinless and spinful multipoles, respectively.
  }
  \label{tab_am}
  \centering
  \begin{tabular}{ccccccc}
  \hline\hline
  rank & type & irrep.   & symbol               & expression \\ 
  \hline
  0    & E    & ${\rm A}_{1g}^{+}$ & $Q^{({\rm a})}_{0}$   & $1$ \\
  \hline
  1    & M    & ${\rm A}_{2g}^{-}$ & $M^{({\rm a})}_{z}$   & $l_{z}$ \\
       & M    & ${\rm E}_{g}^{-}$  & $M^{({\rm a})}_{x}, M^{({\rm a})}_{y}$   & $l_{x}, l_{y}$ \\
  \hline
  2    & E    & ${\rm A}_{1g}^{+}$ & $Q^{({\rm a})}_{u}$   & $3z^{2}-r^{2}$ \\
        & E    & ${\rm B}_{1g}^{+}$ & $Q^{({\rm a})}_{v}$   & $x^{2} - y^{2}$ \\
        & E    & ${\rm E}_{g}^{+}$ &  $Q^{({\rm a})}_{yz}, Q^{({\rm a})}_{zx}$   &  $yz, zx$ \\
        & E   & ${\rm B}_{2g}^{+}$ & $Q^{({\rm a})}_{xy}$   &  $xy$ \\

  \hline
  3    & M    & ${\rm E}_{g}^{-}$  & $M^{({\rm a})}_{3a}, M^{({\rm a})}_{3b}$   & $\left(x^{2}-y^{2}\right) l_{x}-2 x y l_{y}, 2 x y l_{x}+\left(x^{2}-y^{2}\right) l_{y}$ \\
       & M    & ${\rm E}_{g}^{-}$  & $M^{({\rm a})}_{3u}, M^{({\rm a})}_{3v}$   & $\left(5 z^{2}-r^{2}\right) l_{x}+2 x\left(5 z l_{z}-\bm{r} \cdot \bm{l}\right), \left(5 z^{2}-r^{2}\right) l_{y}+2 y\left(5 z l_{z}-\bm{r} \cdot \bm{l}\right)$ \\
  \hline
  4    & E    & ${\rm A}_{1g}^{+}$ & $Q^{({\rm a})}_{4}$   & $x^{4} - 3 x^{2} y^{2} - 3 x^{2} z^{2} + y^{4} - 3 y^{2} z^{2} + z^{4}$ \\
       & E    & ${\rm A}_{1g}^{+}$ & $Q^{({\rm a})}_{4u}$  & $- \left(x^{4} - 12 x^{2} y^{2} + 6 x^{2} z^{2} + y^{4} + 6 y^{2} z^{2} - 2 z^{4}\right)$ \\
       & E    & ${\rm B}_{1g}^{+}$ & $Q^{({\rm a})}_{4v}$  & $  \left(x^{2} - y^{2}\right) \left(x^{2} + y^{2} - 6 z^{2}\right)$ \\
       & E    & ${\rm A}_{2g}^{+}$ & $Q^{\alpha ({\rm a})}_{4z}$  & $ x y \left(x^{2} - y^{2}\right)$ \\
       & E    & ${\rm B}_{2g}^{+}$ & $Q^{\beta ({\rm a})}_{4z}$  & $x y\left(7 z^2-r^2\right)$ \\

& E    & ${\rm E}_{g}^{+}$ & $Q^{\alpha({\rm a})}_{4x}, Q^{\alpha({\rm a})}_{4y}$    &  $yz(y^{2}-z^{2}), zx(z^{2}-x^{2})$ \\
 & E    & ${\rm E}_{g}^{+}$ & $Q^{\beta({\rm a})}_{4x}, Q^{\beta({\rm a})}_{4y}$    &  $yz(7x^{2}-r^{2}), zx(7y^{2}-r^{2})$ \\
  \hline \hline 
  0    & E    & ${\rm A}_{1g}^{+}$ & $Q'^{({\rm a})}_{0}$  & $\bm{l}\cdot\bm{\sigma}$ \\
  \hline
  1    & ET   & ${\rm A}_{2g}^{+}$ & $G^{({\rm a})}_{z}$   & $l_{x} \sigma_{y} - l_{y} \sigma_{x}$ \\
  \hline
  3    & ET   & ${\rm A}_{2g}^{+}$ & $G^{\alpha ({\rm a})}_{z}$     & $M^{({\rm a})}_{3u} \sigma_{y}-M^{({\rm a})}_{3v} \sigma_{x}$ \\
    \hline
  4    & E    & ${\rm A}_{2g}^{+}$ & $Q'^{\alpha ({\rm a})}_{4z}$   & $M^{({\rm a})}_{3a} \sigma_{y} + M^{({\rm a})}_{3b} \sigma_{x}$ \\
  \hline\hline
  \end{tabular}
\end{table*}

Let us first introduce the atomic multipole basis within the spinful $d$ orbitals.
There are 25 (75) independent spinless (spinful) atomic multipoles.
Here, we only show the multipoles appeared in the Hamiltonian and some relevant ones discussed in the main text, $Q_{4z}^{\alpha}$, $Q'^{\alpha}_{4z}$, $G_{z}$, and $G_{z}^{\alpha}$.
The explicit expressions of these multipoles are summarized in Table~\ref{tab_am}, and their matrix elements in the five $d$ orbitals ($d_{u}, d_{v}, d_{yz}, d_{zx}, d_{xy}$) are given by
\begin{widetext}
\begin{align}
& Q^{({\rm a})}_{0} =
\frac{1}{\sqrt{10}}
  \begin{pmatrix}
    1 & 0 & 0 & 0 & 0 \\
    0 & 1 & 0 & 0 & 0 \\
    0 & 0 & 1 & 0 & 0 \\
    0 & 0 & 0 & 1 & 0 \\
    0 & 0 & 0 & 0 & 1 
  \end{pmatrix} \sigma_{0}, \cr
& M^{({\rm a})}_{z} =
\frac{\sqrt{5}}{10}
\begin{pmatrix}
0 & 0 & 0 & 0 & 0 \\
0 & 0 & 0 & 0 & -2i \\
0 & 0 & 0 & i & 0 \\
0 & 0 & -i & 0 & 0 \\
0 & 2i & 0 & 0 & 0
\end{pmatrix} \sigma_{0}, \quad 
M^{({\rm a})}_{x} =
\frac{\sqrt{5}}{10}
\begin{pmatrix}
  0 & 0 & \sqrt{3}i & 0 & 0 \\
  0 & 0 & i & 0 & 0 \\
  -\sqrt{3}i & -i & 0 & 0 & 0 \\
  0 & 0 & 0 & 0 & i \\
  0 & 0 & 0 & -i & 0
\end{pmatrix}  \sigma_{0}, \quad 
M^{({\rm a})}_{y} =
\frac{\sqrt{5}}{10} 
\begin{pmatrix}
  0 & 0 & 0 & -\sqrt{3}i & 0 \\
  0 & 0 & 0 & i & 0 \\
  0 & 0 & 0 & 0 & -i \\
  \sqrt{3}i & -i & 0 & 0 & 0 \\
  0 & 0 & i & 0 & 0
\end{pmatrix}  \sigma_{0}, \quad 
\cr
& Q^{({\rm a})}_{u} = 
\frac{\sqrt{7}}{14}
\begin{pmatrix}
2 & 0 & 0 & 0 & 0 \\
0 & -2 & 0 & 0 & 0 \\
0 & 0 & 1 & 0 & 0 \\
0 & 0 & 0 & 1 & 0 \\
0 & 0 & 0 & 0 & -2
\end{pmatrix}  \sigma_{0}, \quad
Q^{({\rm a})}_{v} = 
\frac{\sqrt{7}}{14}
\begin{pmatrix}
0 & -2 & 0 & 0 & 0 \\
-2 & 0 & 0 & 0 & 0 \\
0 & 0 & -\sqrt{3} & 0 & 0 \\
0 & 0 & 0 & \sqrt{3} & 0 \\
0 & 0 & 0 & 0 & 0
\end{pmatrix}  \sigma_{0}, \cr
& 
Q^{({\rm a})}_{yz} = 
\frac{\sqrt{7}}{14}
\begin{pmatrix}
0 & 0 & 1 & 0 & 0 \\
0 & 0 & -\sqrt{3} & 0 & 0 \\
1 & -\sqrt{3} & 0 & 0 & 0 \\
0 & 0 & 0 & 0 & \sqrt{3} \\
0 & 0 & 0 & \sqrt{3} & 0
\end{pmatrix}  \sigma_{0}, \quad
Q^{({\rm a})}_{zx} = 
\frac{\sqrt{7}}{14}
\begin{pmatrix}
0 & 0 & 0 & 1 & 0 \\
0 & 0 & 0 & \sqrt{3} & 0 \\
0 & 0 & 0 & 0 & \sqrt{3}  \\
1 & \sqrt{3}  & 0 & 0 & 0 \\
0 & 0 & \sqrt{3}  & 0 & 0
\end{pmatrix}  \sigma_{0}, \quad
Q^{({\rm a})}_{xy} = 
\frac{\sqrt{7}}{14}
\begin{pmatrix}
0 & 0 & 0 & 0 & -2 \\
0 & 0 & 0 & 0 & 0 \\
0 & 0 & 0 & \sqrt{3} & 0 \\
0 & 0 & \sqrt{3}  & 0 & 0 \\
-2 & 0 & 0 & 0 & 0
\end{pmatrix}  \sigma_{0},
 \cr
& M^{({\rm a})}_{3a} = 
\frac{\sqrt{2}}{4}
\begin{pmatrix}
0 & 0 & 0 & 0 & 0 \\
0 & 0 & -i & 0 & 0 \\
0 & i & 0 & 0 & 0 \\
0 & 0 & 0 & 0 & i \\
0 & 0 & 0 & -i & 0
\end{pmatrix}  \sigma_{0}, \quad
M^{({\rm a})}_{3b} = 
\frac{\sqrt{2}}{4}
\begin{pmatrix}
  0 & 0 & 0 & 0 & 0 \\
  0 & 0 & 0 & i & 0 \\
  0 & 0 & 0 & 0 & i \\
  0 & -i & 0 & 0 & 0 \\
  0 & 0 & -i & 0 & 0
\end{pmatrix}  \sigma_{0}, \cr
& 
M^{({\rm a})}_{3u} = 
\frac{\sqrt{10}}{20}
\begin{pmatrix}
0 & 0 & 2i & 0 & 0 \\
0 & 0 & -\sqrt{3}i & 0 & 0 \\
-2i & \sqrt{3}i & 0 & 0 & 0 \\
0 & 0 & 0 & 0 & -\sqrt{3}i \\
0 & 0 & 0 & \sqrt{3}i & 0
\end{pmatrix} \sigma_{0}, \quad
M^{({\rm a})}_{3v} = 
\frac{\sqrt{10}}{20}
\begin{pmatrix}
0 & 0 & 0 & -2i & 0 \\
0 & 0 & 0 & -\sqrt{3}i & 0 \\
0 & 0 & 0 & 0 & \sqrt{3}i \\
2i & \sqrt{3}i & 0 & 0 & 0 \\
0 & 0 & -\sqrt{3}i & 0 & 0
\end{pmatrix}  \sigma_{0}, \cr
& Q^{({\rm a})}_{4} =
\frac{\sqrt{15}}{30}
\begin{pmatrix}
  3 & 0 & 0 & 0 & 0 \\
  0 & 3 & 0 & 0 & 0 \\
  0 & 0 & -2 & 0 & 0 \\
  0 & 0 & 0 & -2 & 0 \\
  0 & 0 & 0 & 0 & -2
  \end{pmatrix}  \sigma_{0}, \quad
Q^{({\rm a})}_{4u} =
\frac{\sqrt{21}}{42}
\begin{pmatrix}
3 & 0 & 0 & 0 & 0 \\
0 & -3 & 0 & 0 & 0 \\
0 & 0 & -2& 0 & 0 \\
0 & 0 & 0 & -2& 0 \\
0 & 0 & 0 & 0 & 4
\end{pmatrix}  \sigma_{0}, \quad
Q^{({\rm a})}_{4v} =
\frac{\sqrt{7}}{14}
\begin{pmatrix}
0 & -\sqrt{3} & 0 & 0 & 0 \\
-\sqrt{3} & 0 & 0 & 0 & 0 \\
0 & 0 & 2 & 0 & 0 \\
0 & 0 & 0 & -2 & 0 \\
0 & 0 & 0 & 0 & 0
\end{pmatrix}  \sigma_{0}, \cr
& Q^{\alpha ({\rm a})}_{4z} =
\frac{1}{2}
\begin{pmatrix}
0 & 0 & 0 & 0 & 0 \\
0 & 0 & 0 & 0 & 1 \\
0 & 0 & 0 & 0 & 0 \\
0 & 0 & 0 & 0 & 0 \\
0 & 1 & 0 & 0 & 0
\end{pmatrix}  \sigma_{0}, \cr
& Q'^{({\rm a})}_{0} = \frac{1}{\sqrt{3}} \bm{M}^{({\rm a})} \cdot \bm{\sigma}, \cr
& G^{({\rm a})}_{z} =  \frac{1}{\sqrt{2}} \left(M^{({\rm a})}_{x} \sigma_{y} - M^{({\rm a})}_{y} \sigma_{x}\right), \quad
G^{\alpha ({\rm a})}_{z} = \frac{1}{\sqrt{2}} \left(M^{({\rm a})}_{3u} \sigma_{y}-M^{({\rm a})}_{3v} \sigma_{x}\right), \quad
Q'^{\alpha ({\rm a})}_{4z} = \frac{1}{\sqrt{2}} \left(M^{({\rm a})}_{3a} \sigma_{y}+M^{({\rm a})}_{3b} \sigma_{x}\right).
\label{eq_am}
\end{align}
\end{widetext}
where the superscript (a) denotes the atomic multipole, $\sigma_{i}\, (i=x,y,z)$ and $\sigma_{0}$ are the Pauli matrices and $2\times2$ identity matrix in spin space.
In this Appendix, these atomic multipoles are normalized as $\mathrm{Tr} (X^{({\rm a})}_{i} Y^{({\rm a})}_{j}) = \delta_{XY}\delta_{ij}, \, (X,Y = Q/M/G)$.

\subsubsection{Momentum multipole basis \label{sec:mm}}

To decompose the real and imaginary parts of the nearest-neighbor hopping in the $xy$ plane, we introduce the electric and magnetic toroidal momentum multipole basis as
\begin{align}
     & {\rm A}_{1g}^{+} : Q_{0}^{(1)} (\bm{k}) = \cos(k_{x}) + \cos(k_{y}), \\
     & {\rm B}_{1g}^{+} : Q_{v}^{(1)} (\bm{k}) = \cos(k_{x}) - \cos(k_{y}), \\
     & {\rm E}_{u}^{-} : T_{x}^{(1)} (\bm{k}) = \sqrt{2}\sin(k_{x}), \quad T_{y}^{(1)} (\bm{k}) = \sqrt{2}\sin(k_{y}),
\end{align}
where the superscript (1) denotes the nearest-neighbor hopping within the $xy$ plane. 

Similarly, the electric and magnetic toroidal momentum multipole basis for the nearest-neighbor hoppings along the $z$ direction are given by
\begin{align}
     & {\rm A}_{1g}^{+} : Q_{0}^{(2)} (\bm{k}) = \sqrt{2} \cos(k_{z}), \\
     & {\rm A}_{2u}^{-} :  T_{z}^{(2)} (\bm{k}) = \sqrt{2} \sin(k_{z}),
\end{align}
and those for the next-nearest-neighbor hoppings along the $z$ direction are given by
\begin{align}
     & {\rm A}_{1g}^{+} : Q_{0}^{(3)} (\bm{k}) = \sqrt{2} \left[ \cos(k_{x}) + \cos(k_{y}) \right] \cos(k_{z}), \\
     & {\rm B}_{1g}^{+} : Q_{v}^{(3)} (\bm{k}) = \sqrt{2} \left[ \cos(k_{x}) - \cos(k_{y}) \right] \cos(k_{z}), \\
     & {\rm E}_{g}^{+} : Q_{yz}^{(3)} (\bm{k}) = 2\sin(k_{y})\sin(k_{z}), \quad Q_{zx}^{(3)} (\bm{k}) = 2\sin(k_{z})\sin(k_{x}), \\
     & {\rm A}_{2u}^{-} :  T_{z}^{(3)} (\bm{k}) = \sqrt{2} \left[ \cos(k_{x}) + \cos(k_{y}) \right]\sin(k_{z}), \\
     & {\rm B}_{2u}^{-} : T_{3v}^{(3)} (\bm{k}) = \sqrt{2} \left[ \cos(k_{x}) - \cos(k_{y}) \right] \sin(k_{z}), \\
     & {\rm E}_{u}^{-} : T_{x}^{(3)} (\bm{k}) = 2\sin(k_{x})\cos(k_{z}), \quad T_{y}^{(3)} (\bm{k}) = 2\sin(k_{y})\cos(k_{z}),
\end{align}
where the superscripts (2) and (3) represent the nearest-neighbor and next-nearest-neighbor hoppings along the $z$ direction.  
The momentum multipoles are normalized as $\sum_{\bm{k}} \mathrm{Tr}[X_{i}^{(n)} (\bm{k}) Y_{j}^{(n)} (\bm{k})] = \delta_{XY}\delta_{ij}, \, (X,Y = Q/T, n = 1,2,3)$.

\subsubsection{Multipole decomposition of the tight-binding Hamiltonian \label{sec:md_tbh}}

Next, we express the model Hamiltonian by using the atomic and momentum multipoles given in Secs.~\ref{sec:am} and \ref{sec:mm}, respectively~\cite{hayami2019momentum, Hayami_PhysRevB.101.220403, Hayami_PhysRevB.102.144441, Kusunose_PhysRevB.107.195118}.
Through this procedure, we can clarify the microscopic multipole degrees of freedom in the present system.

Let us consider the total Hamiltonian given in the main text, $h(\bm{k}) \equiv \mathcal{H}_{t}+\mathcal{H}_{\rm SOC} +\mathcal{H}_{\rm CEF}  + \mathcal{H}_{V}$.
The crystalline electric field $\mathcal{H}_{\rm CEF}$ is expressed by using the spinless atomic multipoles belonging to the ${\rm A}_{1g}^{+}$ irreducible representation as
\begin{align}
    \mathcal{H}_{\rm CEF} = \epsilon_{1} Q^{({\rm a})}_{u} + \epsilon_{2} Q^{({\rm a})}_{4} + \epsilon_{3} Q^{({\rm a})}_{4u}.
\end{align}
By using $\Delta_{1}$, $\Delta_{2}$, and $\Delta_{3}$ defined in the main text, $\epsilon_{1}$, $\epsilon_{2}$, and $\epsilon_{3}$ are expressed as follows:
\begin{align}
    \epsilon_{1}   &= \frac{1}{\sqrt{7}} \left(-3\Delta_{1}-3\Delta_{2}+\Delta_{3}\right), \cr
    \epsilon_{2}  &= \frac{\sqrt{15}}{3} \left(\Delta_{2}+\Delta_{3}\right), \cr
    \epsilon_{3} &= \frac{1}{\sqrt{21}} \left(6\Delta_{1}-\Delta_{2}+5\Delta_{3}\right).
\end{align}
The atomic spin--orbit coupling, $\mathcal{H}_{\rm SOC} = \lambda \bm{l} \cdot \bm{\sigma}$, is represented by using the spinful atomic monopole $Q'^{({\rm a})}_{0}$ in ${\rm A}_{1g}^{+}$ as
\begin{align}
    \mathcal{H}_{\rm SOC} = 2\sqrt{15}\lambda Q'^{({\rm a})}_{0}.
\end{align}
Similarly, the symmetry-breaking term $\mathcal{H}_{V}$ in $D_{\rm 4h}$ is represented by using the spinless atomic hexadecapole $Q^{\alpha ({\rm a})}_{4z}$ belonging to ${\rm A}_{2g}^{+}$ irreducible representation,
\begin{align}
    \mathcal{H}_{V} = 2 V Q^{\alpha ({\rm a})}_{4z}.
\end{align}

Since the kinetic energy term, $\mathcal{H}_{t} = \mathcal{H}_{t}^{(1)} + \mathcal{H}_{t}^{(2)} + \mathcal{H}_{t}^{(3)}$, is also fully symmetric for all the symmetry operations in the point group $D_{\rm 4h}$, only the independent products of the atomic and momentum multipoles belonging to ${\rm A}_{1g}^{+}$ irreducible representation contribute to $\mathcal{H}_{t}$.
Considering ${\rm A}_{1g}^{\pm} \otimes {\rm A}_{1g}^{\pm} = {\rm A}_{2g}^{\pm} \otimes {\rm A}_{2g}^{\pm} = {\rm A}_{1g}^{+}$, we obtain
\begin{widetext}
\begin{align}
    &\mathcal{H}_{\rm t}^{(1)} = 
    t_{1} Q^{({\rm a})}_{0} Q_{0}^{(1)} (\bm{k}) 
    + t_{2} Q^{({\rm a})}_{u} Q_{0}^{(1)} (\bm{k}) 
    + t_{3} Q^{({\rm a})}_{v} Q_{v}^{(1)} (\bm{k})  
    + t_{4} Q^{({\rm a})}_{4} Q_{0}^{(1)} (\bm{k}) 
    + t_{5} Q^{({\rm a})}_{4u} Q_{0}^{(1)} (\bm{k})
    + t_{6} Q^{({\rm a})}_{4v} Q_{v}^{(1)} (\bm{k}),
    \cr &
    \mathcal{H}_{\rm t}^{(2)} = 
    t_{7} Q^{({\rm a})}_{0} Q_{0}^{(2)} (\bm{k}) 
    + t_{8} Q^{({\rm a})}_{u} Q_{0}^{(2)} (\bm{k}) 
    + t_{9} Q^{({\rm a})}_{4} Q_{0}^{(2)} (\bm{k})  
    + t_{10} Q^{({\rm a})}_{4u} Q_{0}^{(2)} (\bm{k}),
    \cr & 
    \mathcal{H}_{\rm t}^{(3)} =
       t_{11} Q^{({\rm a})}_{0} Q_{0}^{(3)} (\bm{k}) 
    + t_{12} Q^{({\rm a})}_{u} Q_{0}^{(3)} (\bm{k}) 
    + t_{13} Q^{({\rm a})}_{v} Q_{v}^{(3)} (\bm{k})  
    + t_{14} \frac{1}{\sqrt{2}} \left(
    Q^{({\rm a})}_{yz} Q_{yz}^{(3)} (\bm{k}) 
    +
    Q^{({\rm a})}_{zx} Q_{zx}^{(3)} (\bm{k}) 
    \right)
    + t_{15} Q^{({\rm a})}_{4} Q_{0}^{(3)} (\bm{k})
    + t_{16} Q^{({\rm a})}_{4u} Q_{0}^{(3)} (\bm{k})
    \cr & \qquad \quad
    + t_{17} Q^{({\rm a})}_{4v} Q_{v}^{(3)} (\bm{k})
    + t_{18} \frac{1}{\sqrt{2}} \left(
    Q^{\alpha({\rm a})}_{4x} Q_{yz}^{(3)} (\bm{k}) 
    +
    Q^{\alpha({\rm a})}_{zx} Q_{zx}^{(3)} (\bm{k}) 
    \right)
    + t_{19} \frac{1}{\sqrt{2}} \left(
    Q^{\beta({\rm a})}_{4x} Q_{yz}^{(3)} (\bm{k}) 
    +
    Q^{\beta({\rm a})}_{zx} Q_{zx}^{(3)} (\bm{k}) 
    \right)  .
\end{align}
\end{widetext}
By using the Slater-Koster parameters given in the main text, $t_{1\sigma}$, $t_{1\pi}$, $t_{z 1 \sigma}$, $t_{z 1 \pi}$,  $t_{z 2 \sigma}$, $t_{z 2 \pi}$, $t_{i}\, (i = 1 \sim 19)$ are represented by
\begin{align}
     t_{1}&=\frac{4\sqrt{5}}{5} \left(2 t_{1\sigma} + t_{1\pi}\right), \cr 
     t_{2}&=-\frac{\sqrt{14}}{7}\left(4 t_{1\sigma} + t_{1\pi}\right), \cr 
     t_{3}&=-\frac{\sqrt{42}}{7}\left(4 t_{1\sigma} + t_{1\pi}\right),   \cr
     t_{4}&=\frac{4 \sqrt{30}}{15}\left(3t_{1\sigma} - t_{1\pi}\right), \cr 
     t_{5}&=-\frac{2 \sqrt{42}}{21}\left(3t_{1\sigma} - t_{1\pi}\right), \cr 
     t_{6}&=-\frac{2\sqrt{14}}{7}\left(3 t_{1\sigma} - t_{1\pi}\right).
\end{align}

\begin{align}
t_{7} & = \frac{\sqrt{10}}{5}(2t_{z 1 \pi} + t_{z 1 \sigma}), \cr
t_{8} & = 2\frac{\sqrt{7}}{7}(t_{z 1 \pi} + t_{z 1 \sigma}), \cr
t_{9} & = -\frac{\sqrt{15}}{15}(4t_{z 1 \pi} - 3t_{z 1 \sigma}), \cr
t_{10}& = -\frac{\sqrt{21}}{21}(4t_{z 1 \pi} - 3t_{z 1 \sigma}).
\end{align}

\begin{align}
t_{11} &= \frac{2\sqrt{10}}{5}(2t_{z 2 \pi} + t_{z 2 \sigma}), \cr
t_{12} &= \frac{\sqrt{7}}{7}(t_{z 2 \pi} + t_{z 2 \sigma}), \cr 
t_{13} &= -\frac{\sqrt{21}}{7}(t_{z 2 \pi} + t_{z 2 \sigma}), \cr
t_{14} &= \frac{2\sqrt{21}}{7}(t_{z 2 \pi} + t_{z 2 \sigma}), \cr
t_{15} &= \frac{\sqrt{15}}{30}(4t_{z 2 \pi} - 3t_{z 2 \sigma}), \cr
t_{16} &= \frac{5\sqrt{21}}{84}(4t_{z 2 \pi} - 3t_{z 2 \sigma}), \cr
t_{17} &=  -\frac{5\sqrt{7}}{28}(4t_{z 2 \pi} - 3t_{z 2 \sigma}), \cr
t_{18} &= 0, \cr
t_{19} &= \frac{\sqrt{7}}{7}(4t_{z 2 \pi} - 3t_{z 2 \sigma}).
\end{align}

Notably, although there are 19 independent parameters allowed by the symmetry, only the six independent Slater-Koster parameters appear in the hopping Hamiltonian. 
This is because the Slater-Koster approach assumes the axial symmetry along the bond direction and neglects the surrounding environment of the actual system.

\subsection{Essential model parameters in response tensors \label{sec:extract_parameter}}

In this section, we show the essential model parameters to give a nonzero thermal average of the electric toroidal dipole under the external magnetic field along the $x$ direction $H_{x} \neq 0$, and the linear electric conductivity tensors under the external magnetic fields, by using the systematic analysis method given in Refs.~\cite{Hayami_PhysRevB.102.144441} and \cite{Oiwa_doi:10.7566/JPSJ.91.014701}, which has been used to analyze the essential model parameters of nonlinear (spin) transport~\cite{Oiwa_doi:10.7566/JPSJ.91.014701, Yatsushiro_PhysRevB.105.155157, hayami2022spinconductivity, Hayami_PhysRevB.106.024405, Hayami_PhysRevB.106.014420, Kirikoshi_PhysRevB.107.155109} and nonreciprocal magnon dispersion~\cite{Hayami_PhysRevB.105.014404}.

\subsubsection{Thermal average of the electric toroidal dipole}

Let us begin with the essential model parameters for the thermal average of the electric toroidal dipole $G^{({\rm a})}_{z}$ given in Eq.~(8) in the main text.
The essential model parameters for $\braket{G^{({\rm a})}_{z}} \neq 0$ under $H_{x} \neq 0$ are extracted by analytically evaluating the low-order contributions of the following quantity~\cite{Hayami_PhysRevB.102.144441, Oiwa_doi:10.7566/JPSJ.91.014701},
\begin{align}
    \Gamma^{i}(G^{({\rm a})}_{z}) = \sum_{\bm{k}} \mathrm{Tr}\left[G^{({\rm a})}_{z} h^{i}(\bm{k})\right],
    \label{eq:g1_Gza}
\end{align}
where $h^{i}(\bm{k})$ denotes the $i$-th power of the Hamiltonian matrix including the Zeeman coupling with $H_{x} \neq 0$ at wave vector $\bm{k}$.
The summation of the momentum $\bm{k}$ is taken over 10$^{3}$ grid.

The lowest- and next-lowest-order contributions $i = 5$ and $i = 6$ are explicitly given by
\begin{align}
\Gamma^{5}(G^{({\rm a})}_{z}) &= -120 V H_{x}^{4} , 
\label{eq:Gz_5} \\
\Gamma^{6}(G^{({\rm a})}_{z}) &= 
- 6 V \lambda \cr &\times \left[100 H_{x}^{4} + \left(- 8 t_{1\pi} t_{1\sigma} - t_{z2\pi} t_{z2\sigma} - 32 t_{1\sigma}^{2} - t_{z2\sigma}^{2}\right) H_{x}^{2} \right] . \cr
\label{eq:Gz_6}
\end{align}
As a result, $V$ is the essential model parameter for $\braket{G^{({\rm a})}_{z}} \neq 0$ under $H_{x} \neq 0$.
Notably, the second term in Eq.~(\ref{eq:Gz_6}) indicates that the coupling between $V$ and $\lambda$ gives rise to the additional contributions proportional to $H_{x}^{2}$ in $\braket{G^{({\rm a})}_{z}}$.
These results are consistent with the numerical calculation results shown in Figs.~\ref{Fig: MF_Gz}(a) and \ref{Fig: MF_Gz}(b) in the main text.

\subsubsection{Unconventional Hall conductivity tensors}

We also identify the essential model parameters in the unconventional Hall conductivity tensors $\sigma_{\mu\nu}^{\rm H}(H_x)$ given in Eq.~(11) in the main text by evaluating the following quantity~\cite{Oiwa_doi:10.7566/JPSJ.91.014701},
\begin{align}
    \Gamma_{\mu\nu}^{ij} (H_{\eta}) = \sum_{\bm{k}} \mathrm{Tr}\left[J_{\mu \bm{k}} h^{i}(\bm{k}) J_{\nu \bm{k}}^{} h^{j}(\bm{k})\right],
    \label{eq:cond}
\end{align}
where $J_{\mu \bm{k}}$ is the $\mu$ directional electric current operators at $\bm{k}$ and $h(\bm{k})$ includes the Zeeman coupling with $H_{\eta} \neq 0$.
The summation of the momentum $\bm{k}$ is taken over 10$^{3}$ grid.
Since the essential model parameters are contained in any pairs of $(i,j)$ in Eq.~(\ref{eq:cond}), we only show several lowest-order contributions to Eq.~(\ref{eq:cond}).

First, we focus on $\Gamma_{zx}^{ij}(H_{x})$ that corresponds to the unconventional Hall conductivity tensor $\sigma_{zx}^{\rm H}(H_x)$.
The lowest-order contributions to $\mathrm{Im}\left[\Gamma_{zx}^{ij}(H_{x})\right] = -\mathrm{Im}\left[\Gamma_{xz}^{ij}(H_{x})\right]$ that proportional to $H_{x}$ and $H_{x}^{3}$ are explicitly given by
\begin{widetext}
\begin{align}
\mathrm{Im}\left[\Gamma_{zx}^{03}(H_{x})\right]
&=
- \frac{3 \sqrt{2}}{10} V H_{x} \left(8 t_{1\pi} t_{z1\pi} t_{z2\pi} - 3 t_{1\pi} t_{z1\pi} t_{z2\sigma} - 12 t_{1\sigma} t_{z1\pi} t_{z2\sigma} - 12 t_{1\sigma} t_{z1\sigma} t_{z2\pi} - 6 t_{1\sigma} t_{z1\sigma} t_{z2\sigma}\right), \\
\mathrm{Im}\left[\Gamma_{zx}^{05}(H_{x})\right]
&=
- \frac{3 \sqrt{2}}{10} V H_{x}^{3} \left(148 t_{1\pi} t_{z1\pi} t_{z2\pi} - 36 t_{1\pi} t_{z1\pi} t_{z2\sigma} + 12 t_{1\pi} t_{z1\sigma} t_{z2\pi} + 9 t_{1\pi} t_{z1\sigma} t_{z2\sigma} - 138 t_{1\sigma} t_{z1\pi} t_{z2\sigma} - 180 t_{1\sigma} t_{z1\sigma} t_{z2\pi} - 72 t_{1\sigma} t_{z1\sigma} t_{z2\sigma}\right).
\end{align}
We find that all the terms in $\mathrm{Im}\left[\Gamma_{zx}^{ij}(H_{x})\right]$ are proportional to $V H_{x}^{2m+1}$ and hence $\sigma_{zx}^{\rm H}(H_x)$ is an odd function of $H_{x}$:
\begin{align}
\sigma_{zx}^{\rm H}(H_x) = V \sum_{m = 0} H_{x}^{2m+1}  F_{m}(V,\lambda,\Delta_{1},\Delta_{2},\Delta_{3},t_{1\pi},t_{1\sigma},t_{z1\pi}, t_{z1\sigma},t_{z2\pi}, t_{z2\sigma}),
\label{eq:cond_zxx}
\end{align}
\end{widetext}
where $F_{m}$ is a function of the parameters.
Therefore, the essential model parameters for $\sigma_{zx}^{\rm H}(H_x) \neq 0$ is $V$ under $H_{x} \neq 0$.
These results are consistent with the numerical calculation results shown in Figs.~\ref{Fig: PHall_Hdep}(a) and \ref{Fig: PHall_Hdep}(b) in the main text.

On the other hand, the lowest-order contributions to $\mathrm{Im}\left[\Gamma_{yz}^{ij}(H_{x})\right] = -\mathrm{Im}\left[\Gamma_{zy}^{ij}(H_{x})\right]$ that corresponds to the conventional Hall conductivity tensor $\sigma_{yz}^{\rm H}(H_x)$ are explicitly given by
\begin{widetext}
\begin{align}
\mathrm{Im}\left[\Gamma_{yz}^{02}(H_{x})\right]
&= 
- \frac{3 \sqrt{2}}{5} H_{x} \left(4 t_{1\pi} t_{z1\pi} t_{z2\pi} + 3 t_{1\pi} t_{z1\sigma} t_{z2\sigma} + 12 t_{1\sigma} t_{z1\pi} t_{z2\sigma} + 12 t_{1\sigma} t_{z1\sigma} t_{z2\sigma}\right),
\\
\mathrm{Im}\left[\Gamma_{yz}^{04}(H_{x})\right]
&=
- \frac{12 \sqrt{2}}{5} H_{x}^{3} \left(8 t_{1\pi} t_{z1\pi} t_{z2\pi} + 9 t_{1\pi} t_{z1\sigma} t_{z2\sigma} + 36 t_{1\sigma} t_{z1\pi} t_{z2\sigma} + 39 t_{1\sigma} t_{z1\sigma} t_{z2\sigma}\right).
\end{align}
\end{widetext}
We find that all the terms in $\mathrm{Im}\left[\Gamma_{yz}^{ij}(H_{x})\right]$ are proportional to the odd power of $H_{x}$.
Unlike the unconventional Hall conductivity tensor $\sigma_{zx}^{\rm H}(H_x)$, $V$ is not essential for the conventional Hall conductivity tensor $\sigma_{yz}^{\rm H}(H_x) \neq 0$.
These results are consistent with the numerical calculation results shown in Figs.~\ref{Fig: NHall_Hdep}(a) and \ref{Fig: NHall_Hdep}(b) in the main text.

\subsubsection{Magnetoconductivity tensors}

We here show the essential model parameters in the magnetoconductivity tensors $\sigma_{\mu\nu}^{\rm MC}(H)$ for $H_\eta=H_\gamma=H$ given in Eq.~(15) in the main text by evaluating Eq.~(\ref{eq:cond}).

First, we focus on $\Gamma_{xy}^{ij} (H_{x})$ that corresponds to the magnetoconductivity tensor $\sigma_{xy}^{\rm MC}(H_x)$ by setting $\eta=\gamma=x$.
The lowest-order contributions to $\mathrm{Re}\left[\Gamma_{xy}^{ij}(H_{x})\right] = \mathrm{Re}\left[\Gamma_{yx}^{ij}(H_{x})\right]$ that proportional to $H_{x}^{2}$ and $H_{x}^{4}$ are explicitly given by
\begin{widetext}
\begin{align}
\mathrm{Re}\left[\Gamma_{xy}^{05}(H_{x})\right]
&=
- \frac{1}{512} V H_{x}^{2} \left(2560 t_{1\pi}^{2} t_{z2\pi}^{2} - 1536 t_{1\pi}^{2} t_{z2\pi} t_{z2\sigma} - 192 t_{1\pi}^{2} t_{z2\sigma}^{2} + 6144 t_{1\pi} t_{1\sigma} t_{z2\pi}^{2} - 14592 t_{1\pi} t_{1\sigma} t_{z2\pi} t_{z2\sigma} \right.
 \cr & \qquad\qquad\qquad 
- 8448 t_{1\pi} t_{1\sigma} t_{z2\sigma}^{2} + 6144 t_{1\sigma}^{2} t_{z2\pi}^{2} - 30720 t_{1\sigma}^{2} t_{z2\pi} t_{z2\sigma} - 11520 t_{1\sigma}^{2} t_{z2\sigma}^{2} + 768 t_{z2\pi}^{4} 
\cr & \qquad\qquad\qquad 
\left. - 1728 t_{z2\pi}^{3} t_{z2\sigma} - 4632 t_{z2\pi}^{2} t_{z2\sigma}^{2} - 1644 t_{z2\pi} t_{z2\sigma}^{3} - 135 t_{z2\sigma}^{4}\right),
\\
\mathrm{Re}\left[\Gamma_{xy}^{07}(H_{x})\right]
&=
- \frac{3}{256} V H_{x}^{4} \left(12544 t_{1\pi}^{2} t_{z2\pi}^{2} - 7296 t_{1\pi}^{2} t_{z2\pi} t_{z2\sigma} - 1248 t_{1\pi}^{2} t_{z2\sigma}^{2} + 29696 t_{1\pi} t_{1\sigma} t_{z2\pi}^{2} \right. 
\cr & \qquad\qquad\qquad
- 68224 t_{1\pi} t_{1\sigma} t_{z2\pi} t_{z2\sigma} - 39552 t_{1\pi} t_{1\sigma} t_{z2\sigma}^{2} + 26624 t_{1\sigma}^{2} t_{z2\pi}^{2} - 145408 t_{1\sigma}^{2} t_{z2\pi} t_{z2\sigma}
\cr & \qquad\qquad\qquad
\left. - 52992 t_{1\sigma}^{2} t_{z2\sigma}^{2} + 4544 t_{z2\pi}^{4} - 8640 t_{z2\pi}^{3} t_{z2\sigma} - 23328 t_{z2\pi}^{2} t_{z2\sigma}^{2} - 8256 t_{z2\pi} t_{z2\sigma}^{3} - 747 t_{z2\sigma}^{4}\right).
\end{align}
We find that all the terms in $\mathrm{Re}\left[\Gamma_{xy}^{ij}(H_{x})\right]$ are proportional to $V H_{x}^{2m}$ and hence $\sigma_{xy}^{\rm MC}(H_x)$ is an even function of $H_{x}$:
\begin{align}
\sigma_{xy}^{\rm MC}(H_x) = V \sum_{m = 0} H_{x}^{2m}  F_{m}(V,\lambda,\Delta_{1},\Delta_{2},\Delta_{3},t_{1\pi},t_{1\sigma},t_{z1\pi}, t_{z1\sigma},t_{z2\pi}, t_{z2\sigma}),
\label{eq:cond_xyxx}
\end{align}
\end{widetext}
where $F_{m}$ is a function of the parameters.
Therefore, the essential model parameters for $\sigma_{xy}^{\rm MC}(H_x) \neq 0$ is $V$ under $H_{x} \neq 0$.
These results are consistent with the numerical calculation results shown in Figs.~\ref{Fig: MR1_Hxdep}(a) and \ref{Fig: MR1_Hxdep}(b) in the main text.

Finally, let us discuss the lowest-order contributions to $\mathrm{Re}\left[\Gamma_{zx}^{ij}(H)\right] = \mathrm{Re}\left[\Gamma_{xz}^{ij}(H)\right]$ that corresponds to the magnetoconductivity tensor $\sigma_{zx}^{\rm MC}(H)$ for $H_x=H_z=H$ are explicitly given by
\begin{align}
\mathrm{Re}\left[\Gamma_{zx}^{02}(H) \right]
&=
- \frac{72}{5} H^{2} \left(t_{z2\pi}^{2} + t_{z2\sigma}^{2}\right),
\\
\mathrm{Re}\left[\Gamma_{zx}^{04}(H) \right]
&=
- \frac{144}{5} H^{4} \left(11 t_{z2\pi}^{2} + 10 t_{z2\sigma}^{2}\right).
\end{align}
Unlike the magnetoconductivity tensor $\sigma_{xy}^{\rm MC}(H_x)$, $\sigma_{zx}^{\rm MC}(H) \neq 0$ can be realized without the ferroaxial ordering term $V$.

There are other contributions, $\mathrm{Im}\left[\Gamma_{zx}^{ij}(H)\right] = -\mathrm{Im}\left[\Gamma_{xz}^{ij}(H)\right]$, that corresponds to the unconventional Hall conductivity tensor $\sigma_{zx}^{\rm H}(H_x)$ with $H \neq 0$.
The lowest-order contributions are explicitly given by
\begin{widetext}
\begin{align}
\mathrm{Im}\left[\Gamma_{zx}^{03}(H) \right]
&=
- \frac{3 \sqrt{2}}{10} V H \left(8 t_{1\pi} t_{z1\pi} t_{z2\pi} - 3 t_{1\pi} t_{z1\pi} t_{z2\sigma} - 12 t_{1\sigma} t_{z1\pi} t_{z2\sigma} - 12 t_{1\sigma} t_{z1\sigma} t_{z2\pi} - 6 t_{1\sigma} t_{z1\sigma} t_{z2\sigma}\right),
\\
\mathrm{Im}\left[\Gamma_{zx}^{05}(H) \right]
&=
- \frac{3 \sqrt{2}}{10} V H^{3} \left(
296 t_{1\pi} t_{z1\pi} t_{z2\pi} - 90 t_{1\pi} t_{z1\pi} t_{z2\sigma} + 12 t_{1\pi} t_{z1\sigma} t_{z2\pi} + 9 t_{1\pi} t_{z1\sigma} t_{z2\sigma} + 48 t_{1\sigma} t_{z1\pi} t_{z2\pi} - 366 t_{1\sigma} t_{z1\pi} t_{z2\sigma} \right. 
\cr & \qquad\qquad\qquad \left. 
- 360 t_{1\sigma} t_{z1\sigma} t_{z2\pi} - 162 t_{1\sigma} t_{z1\sigma} t_{z2\sigma}\right).
\end{align}
\end{widetext}
Similar to Eq.~(\ref{eq:cond_zxx}), all the terms in $\mathrm{Im}\left[\Gamma_{zx}^{ij}(H)\right]$ are proportional to $V H^{2m+1}$.

\section{Optical unconventional Hall conductivity}
\label{sec: Optical conductivity}

\begin{figure}[htb!]
\begin{center}
\includegraphics[width=0.8 \hsize]{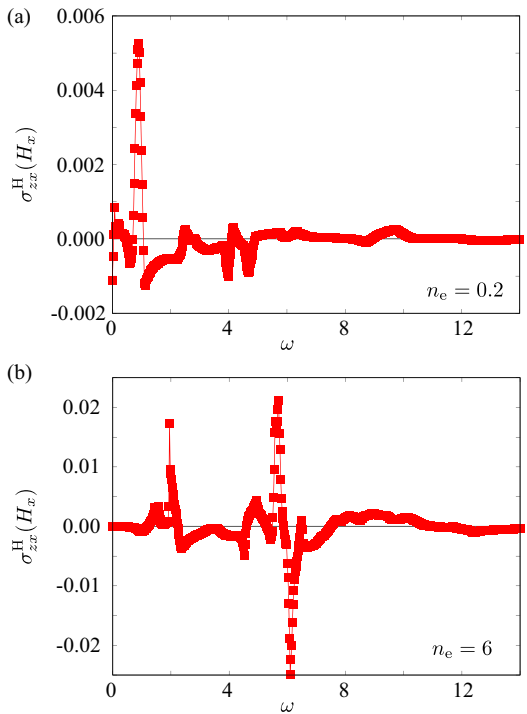}
\caption{
\label{Fig: PHall_omega}
$\omega$ dependence of $\sigma^{\rm H}_{zx}(H_x)$ at $\lambda=2$ for (a) $n_{\rm e}=0.2$ and (b) $n_{\rm e}= 6$. 
}
\end{center}
\end{figure}

We show the optical unconventional Hall conductivity $\sigma^{\rm H}_{zx}(H_x)$ by considering finite frequency $\omega$ in Fig.~\ref{Fig: PHall_omega}. 
The data are calculated at fixed $H_x=0.3$ and $\lambda=2$ for $n_{\rm e}=0.2$ in Fig.~\ref{Fig: PHall_omega}(a) and $n_{\rm e}= 6$ in Fig.~\ref{Fig: PHall_omega}(b).
As shown in Figs.~\ref{Fig: PHall_omega}(a) and \ref{Fig: PHall_omega}(b), $\sigma^{\rm H}_{zx}(H_x)$ becomes nonzero for finite frequency $\omega$, although it shows a complicated behavior including the sign change with changing $\omega$. 
In the insulating case in Fig.~\ref{Fig: PHall_omega}(b), $\sigma^{\rm H}_{zx}(H_x)$ remains zero for a low frequency smaller than the band gap, but it becomes nonzero when the frequency is larger than the band gap. 
Thus, the optical unconventional Hall conductivity is one of the measurements to identify the ferroaxial ordering when the materials are insulating.

\bibliographystyle{apsrev}
\bibliography{ref}

\end{document}